 \documentclass[traditabstract]{aa}
\usepackage{graphicx}
\usepackage{enumerate}
\usepackage{natbib}
\usepackage{txfonts}
\usepackage{verbatim}
\bibpunct{(}{)}{;}{a}{}{,} % to follow the A&A style

\begin{document}

\title{ \mbox{3-D} non-LTE radiative transfer effects in \ion{Fe}{i} lines}
\subtitle{II. Line formation in \mbox{3-D} radiation hydrodynamic simulations}

\author{R.~Holzreuter\inst{1,2}, S.~K.~Solanki\inst{1,3} }

\institute{Max Planck Institute for Solar System Research (MPS), 37191 Katlenburg-Lindau, Germany \and 
 Institute of Astronomy, ETH Zentrum, CH-8092 Zurich, Switzerland \and 
 School of Space Research, Kyung Hee University, Yongin, Gyeonggi 446-701, Republic of Korea\\
 \email{holzreuter@astro.phys.ethz.ch}}

\offprints{R.~Holzreuter}

\date{Received $<$date$>$; accepted $<$date$>$}

%%%%%%%%%%%%%%%%%%%%%%%%%%%%%%%%%%%%%%%%%%%%%%%%%%%%%%%%%%%%%%%%%%%%%%%%
%%%%%%%%%%%%%%%%%%%%%%%%%%%%%%%%%%%%%%%%%%%%%%%%%%%%%%%%%%%%%%%%%%%%%%%%
%\abstract
%{ context }
%{ Aims }
%{ Methods }
%{ Results }
%{ Conclusions %}

\abstract
{ Here we investigate the effects of horizontal radiative transfer (RT) in combination with NLTE on important diagnostic iron lines in a realistic atmosphere. Using a snapshot of a \mbox{3-D} radiation-hydrodynamic (HD) simulation and a multi-level iron atom we compute widely used \ion{Fe}{i} line profiles at three different levels of approximation of the RT (\mbox{3-D} NLTE, \mbox{1-D} NLTE, LTE). By comparing the resulting line profiles and the circumstances of their formation, we gain new insights into the importance of horizontal RT.
We find that the influence of horizontal RT is of the same order of magnitude as that of NLTE, although spatially more localized. Also, depending on the temperature of the surroundings, horizontal RT is found to weaken or strengthen spectral lines. Line depths and equivalent width may differ by up to $20$\% against the corresponding LTE value if \mbox{3-D} RT is applied. Residual intensity contrasts in LTE are found to be larger than those in \mbox{3-D} NLTE by up to a factor of two. When compared to \mbox{1-D} NLTE, we find that horizontal RT weakens the contrast by up to $30$\% almost independently of the angle of line of sight. While the center-to-limb variation (CLV) of the \mbox{1-D} and \mbox{3-D} NLTE contrasts are of similar form, the LTE contrast CLV shows a different run. The determination of temperatures by \mbox{1-D} NLTE inversions of spatially resolved observations may produce errors of up to $200$ K if one neglects \mbox{3-D} RT. We find a linear correlation between the intensity difference of \mbox{1-D} and \mbox{3-D} NLTE and a simple estimate of the temperature in the horizontal environment of the line formation region. This correlation could be used to coarsely correct for the effects of horizontal RT in inversions done in \mbox{1-D} NLTE.
Horizontal RT is less important if one considers spatially averaged line profiles because local line strengthening and weakening occur with similar frequency in our HD atmosphere. Thus, the iron abundance is underestimated by 0.012 dex if calculated using \mbox{1-D} NLTE RT. Since effects of horizontal RT are largest for spatially resolved quantities, the use of \mbox{3-D} RT is particularly important for the interpretation of high spatial resolution observations.

}

\maketitle

%%%%%%%%%%%%%%%%%%%%%%%%%%%%%%%%%%%%%%%%%%%%%%%%%%%%%%%%%%%%%%%%%%%%%%%%
\section{Introduction}\label{sec:hdfe23_intro}
%%%%%%%%%%%%%%%%%%%%%%%%%%%%%%%%%%%%%%%%%%%%%%%%%%%%%%%%%%%%%%%%%%%%%%%%

Due to its countless spectral lines in nearly all wavelength ranges and its relatively high abundance, iron plays an important role in the investigation of solar (and stellar) properties, such as temperature, elemental abundances (iron is often taken as a proxy for all metals), velocity and magnetic fields. The rich spectrum of iron is a result of its rich and complex term structure, which also makes it difficult to achieve the proper degree of realism in the synthesis of its spectral lines. Non-Local Thermodynamic Equilibrium (NLTE) effects have been investigated and identified to be important already a considerable time ago \citep[e.g.,][]{athaylites1972, steenbockholweger1984, boyarchuketal1985, solankisteenbock1988, theveninidiart1999, shchukinatrujillobueno2001}. The most important NLTE effect influencing \ion{Fe}{i} photospheric line formation in the Sun is the so-called UV over-ionization, i.e. the sensitivity of \ion{Fe}{i} to UV irradiation around $300$~nm \citep[see e.g.][ and references therein]{athaylites1972, rutten1988, shchukinatrujillobueno2001} leading to a significant under-population of the \ion{Fe}{i} levels and therefore to a weakening of many diagnostically important lines.

The complex atomic structure forces investigators to make simplifying assumptions, be it in the atomic structure or in the calculation method (e.g. the use of \mbox{1-D} atmospheres or the assumption of LTE) to reduce the problem to a tractable size. A series of works focused on a maximized reproduction of the atomic structure using atoms with hundreds \citep[][]{shchukinatrujillobueno2001} and --- more recently --- up to several thousands \citep[][]{mashonkinaetal2011} of levels and transitions. In addition, the advent of three-dimensional (3-D) hydrodynamic (HD) and magneto-hydrodynamic (MHD) simulations of (magneto-) convection \citep[e.g.][]{nordlund1982, nordlund1983, nordlund1985rev, steinnordlund1998, steinnordlund2001,steinnordlund2001b,voegleretal2004,voegleretal2005} enabled a major increase of the degree of realism of the atmospheric input. \citet{shchukinatrujillobueno2001,shchukinaetal2005} were the first who combined a relatively complete atomic approach including NLTE with a \mbox{3-D} atmospheric input from a HD simulation of \citet{asplundetal2000b,asplundetal2000} to investigate the influence of NLTE effects on the \ion{Fe}{i} spectrum in granular and intergranular regions. However, they too had to make simplifying assumptions: By using a \mbox{1.5-D} approach, they neglected the influence of horizontal radiative transfer (RT).

To date, true \mbox{3-D} RT in combination with NLTE has only been applied to a few atomic species as e.g. by \citet{KiselmanNordlund1995} to \ion{O}{i} lines, \citet{uitenbroek2006} to \ion{Na}{i} and \ion{Ca}{II} IR lines, or, more recently, \citet{leenaartsetal2012} to the chromospheric $H_\alpha$ line, the latter showing a very strong dependance on \mbox{3-D} NLTE. Refer also to \citet{leenaartsetal2012} for more references.  So far, no publication exists which applies true \mbox{3-D} NLTE RT to iron lines even though the influence of horizontal RT in \ion{Fe}{i} lines was impressively demonstrated already some time ago by \citet{stenholmstenflo1977,stenholmstenflo1978} within a simple flux tube model. They showed that enhanced UV over-ionization due to influx of radiation from the hot walls of the flux tube leads to a significant line weakening when compared with \mbox{1-D} NLTE or LTE calculations. \citet{holzreutersolanki2012} (hereafter referred to as Paper~I) refined the investigation of \citet{stenholmstenflo1977,stenholmstenflo1978} by using a more complex atom and a more realistic flux tube geometry \citep[cf.][]{brulsvdluehe2001}. In Paper~I we found qualitatively similar results as \citet{stenholmstenflo1977,stenholmstenflo1978}, although the picture turned out to be more complex and the line weakening less extreme.

In this work, we continue the investigation on the influence of horizontal RT on the solar \ion{Fe}{i} spectrum by enhancing the realism of the atmosphere. Throughout this work, we use a snapshot of a \mbox{3-D} radiation hydrodynamic simulation computed with the MURAM code \citep{voegleretal2005}. Although we do not investigate a magnetic atmosphere, the same effects as described in Paper~I should also exist in a field-free atmosphere, especially at locations with strong horizontal temperature gradients. As mentioned in Paper~I, we encounter not only line weakening by horizontal irradiation from hot environments, but also the inverse effect at locations where the source function of the environment is lower than that of the line-forming region itself. We will show in this work that line strengthening occurs as frequently as line weakening in our HD atmosphere.

Here, we again consider the $524.71 / 525.02$~nm and the $630.15/630.25$~nm line pairs. These four lines are generally used to investigate magnetic features. The $630.15$ and the $630.25$~nm lines have in the past often not been considered ideal for studying convection etc. due to their non-zero Land\'e factors. This view has been mitigated by the success of the Hinode SP \citep{kosugietal2007HINODE,tsunetaetal2008} also for studies unrelated to the magnetic field. Similarly, the IMaX instrument \citep{martinezpilletetal2011} on the Sunrise mission \citep{solankietal2010,Bartholetal2011} has demonstrated the usefulness of the \ion{Fe}{i} $525.02$~nm line also for non-magnetic studies \citep{bellogonzalezetal2010,khomenkoetal2010,rothetal2010}. Another reason we used these lines was to be able to compare with an upcoming investigation based on MHD simulations.

We aim to shed more light on the quantitative differences between \mbox{1-D} NLTE (LTE) and \mbox{3-D} NLTE. We will show that spatially averaged quantities such as the averaged line profile show only a very minor dependence on horizontal RT, but if resolved line profiles are considered, significant differences between \mbox{1-D} and \mbox{3-D} RT line profiles may appear. This is the case for, e.g., RMS intensity contrasts in line cores. It is important to consider spatially resolved line profiles for investigations where high resolution observations are inverted.

%With a similar approach but with the use of LTE only, \citet{fabbianetal2011} demonstrated the influence of Zeeman broadening on
%%%%%%%%%%%%%%%%%%%%%%%%%%%%%%%%%%%%%%%%%%%%%%%%%%%%%%%%%%%%%%%%%%%%%%%%
\section{Model ingredients}\label{sec:hdfe23_model}
%%%%%%%%%%%%%%%%%%%%%%%%%%%%%%%%%%%%%%%%%%%%%%%%%%%%%%%%%%%%%%%%%%%%%%%%

\subsection{Radiative transfer calculations}\label{sec:hdfe23_model_RT}
%%%%%%%%%%%%%%%%%%%%%%%%%%%%%%%%%%%%%%%%%%%%%%%%%%%%%%%%%%%%%%%%%%%%%%%%
All our calculations are done with the \mbox{3-D} RT code RH developed by \citet{uitenbroek2000} and based on the iteration scheme of \citet{rybickihummer1991,rybickihummer1992,rybickihummer1994}. We use a modified version (see Paper~I for details) which uses monotonic parabolic B\'ezier integration of the source function in the integration of the short characteristics \citep{olsonkunasz1987,kunaszauer1988,reesetal1989,socasnavarroetal2000} as proposed by \citet{auer2003}.

\subsection{Atomic model}\label{sec:hdfe23_model_atom}
%%%%%%%%%%%%%%%%%%%%%%%%%%%%%%%%%%%%%%%%%%%%%%%%%%%%%%%%%%%%%%%%%%%%%%%%
\begin{table}
\caption{f-values of our four selected lines}
\begin{center}
\begin{tabular}{llrp{0.25\textwidth}}
 Line [nm] & f-value & source \\
\hline
 524.71 & $2.26\times10^{-06}$ & \citet{blackwelletal1979} \\ 
 525.02 & $1.15\times10^{-05}$ & \citet{blackwelletal1979} \\ 
 630.15 & $3.83\times10^{-02}$ & \citet{bardetal1991} \\
 630.25 & $2.31\times10^{-02}$ & \citet{thevenin1989,thevenin1990}, \newline \\
 & & \citet{socasnavarro2011} \\ 
\hline
\end{tabular}
\end{center}
\label{tab:f-values}
\end{table}
%-------------------------------------------------
We used a similar atomic model ($23$ levels, $33$ lines, using a total of approximately $1300$ wavelength points) as in Paper~I but with the best gf-values available in the literature for the lines used in our investigation (see Table \ref{tab:f-values} for a summary). The atom was tailored to reduce the computational costs but still to represent the most important transitions for our selected lines. A test with a much larger model (121 levels, 357 lines) --- kindly provided by J.\ H.\ M.\ J.\ Bruls --- showed no major difference in the spectrum of the lines discussed here.

The problem of missing iron line opacity in the UV, which may lead to wrong population numbers \citep{brulsetal1992}, was dealt with by artificially increasing the opacities in the relevant wavelength range as proposed by \citet{brulsetal1992}. We refer to Paper~I for a more complete discussion.

The elastic collision rates were calculated according to the semi-classical approach of \citet{ansteeomara1995,barklemomara1997}. For the iron abundance we employed a value of 7.50 \citep{asplundetal2009} for all computations. 

The influence of inelastic collisions with neutral hydrogen, which is still under intense debate \citep[see e.g. the discussion by][ and references therein]{mashonkinaetal2011}, was neglected in our calculations because the formula for the calculation of the rates \citep{drawin1968} is accurate within an order of magnitude only. It seems that --- at least for solar conditions --- the neglect does not have far reaching consequences on NLTE level populations \citep[see e.g.][]{allendeprietoetal2004,mashonkinaetal2011}. However, because these authors used \mbox{1-D} atmospheres for their argumentation, it cannot be excluded completely that in a very inhomogeneous \mbox{3-D} atmosphere, the influence of collisions with neutral hydrogen may at some (mainly cold) locations have some relevance.

\subsection{Model atmospheres}\label{sec:hdfe23_model_atmos}
%%%%%%%%%%%%%%%%%%%%%%%%%%%%%%%%%%%%%%%%%%%%%%%%%%%%%%%%%%%%%%%%%%%%%%%%
\begin{figure*}
\center{ \resizebox{0.85\hsize}{!}{\includegraphics{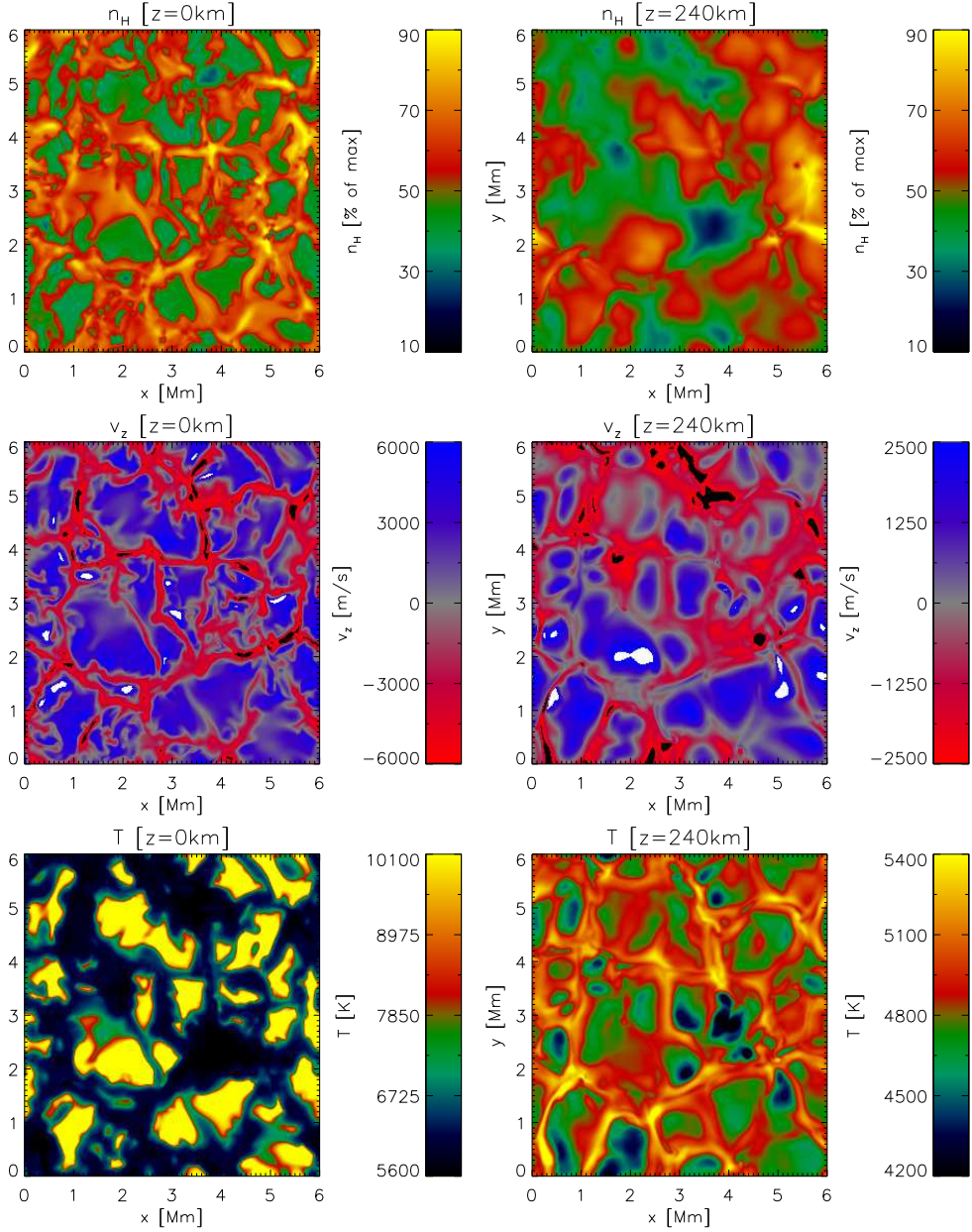}} }
 \caption{ 
 Atmospheric parameters at two heights. Left: $z=0$~km, i.e. at the average level of formation of the continuum. Right: $z=240$~km, approximate height of the line core formation of our three selected lines. From top to bottom the hydrogen number density, $n_H$ (relative to the maximum value at that height), the vertical velocity, $v_z$ (with negative values denoting up-flows), and the temperature T are plotted.
} 
 \label{fig:tv_atmos}
\end{figure*}

The input atmosphere model for our calculations was taken from a realistic \mbox{3-D} radiation hydrodynamic simulation. We used a snapshot from a $B=0$ run of the MURAM code \citep[][]{voegleretal2005} with an open boundary condition at the top of the atmosphere \citep[using the so-called "fiducial layer" method, see e.g.][]{steinnordlund1998} allowing for mass flows into or out of the cube. The original resolution of the cube was $288 \times 288 \times 120$. The completely optically thick parts of the atmosphere sufficiently far below the photosphere (48 pixels) were cut away. The resulting $288 \times 288 \times 72$ cube starts approximately 300 km (average $\tau_{cont}^{500} \approx 200$) below the average $\tau_{cont}^{500}=1$ level and reaches a height of approximately 700 km above it. The spatial extent in $x$ and $y$ direction is $6000$~km each, i.e. the cube has a constant grid spacing of approximately $21\times 21\times 15$~km$^3$. The original Cartesian geometry of the cube was directly employed also in our RT calculations. Due to the large variations of physical quantities across the cube, even a single line may be formed at rather different geometric height ranges at different $(x,y)$-locations. Consequently, an equidistant z-grid has proven to be favorable. 

The temperature structure and the velocity fields were directly taken from the MURAM cube. The H populations ($n_H$) of the ground level were calculated from the MURAM cube density. The population of excited levels of H were neglected. The electron densities $n_e$ were calculated by the RH code in a self consistent way. Fig.~\ref{fig:tv_atmos} shows the spatial distribution of some atmospheric parameters at the average continuum formation level ($z=0$~km; panels on the left) and at a height at which, roughly, the cores of the selected lines are formed ($z=240$~km; panels on the right). Of course, the formation height depends strongly on $x,y$ so that these are just illustrative heights. In Fig.~\ref{fig:tv_atmos} we see the granular structure with hot uprising granules and dense, cold down-flowing intergranular lanes at the $z=0$~km level. At $z=240$~km, the temperature difference between still rising granular and down-flowing intergranular regions is inverted, whereas the density distribution is relatively smooth. The temperature inversion takes place at heights between $z=150$~km and $z=180$~km. 

In order to compare our \mbox{3-D} calculations with \mbox{1-D} computations we constructed plane-parallel atmospheres as in Paper~I by vertically cutting the \mbox{3-D} cube at all $(x,y)$-positions. The corresponding $82944$ \mbox{1-D} problems were solved individually. The resulting \mbox{1-D} population numbers were then re-transferred into a \mbox{3-D} cube for the final calculation of the emerging spectra \mbox{(1.5-D)}. Throughout this work we will use the term \mbox{1-D} for the 1.5-D approach.

\begin{figure*}
\center{ \resizebox{0.7\hsize}{!}{\includegraphics{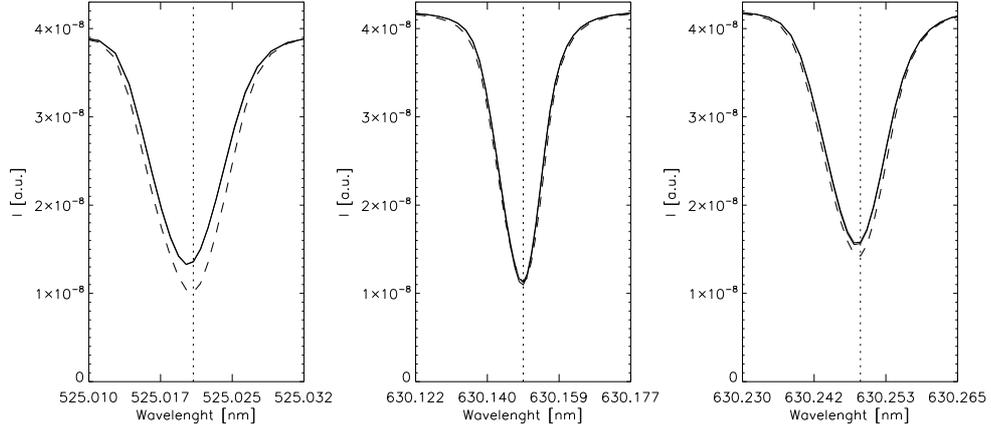}} }
 \caption{ 
 Mean profiles of our three selected lines averaged over the whole $x-y$ plane for three calculations methods (thick solid: \mbox{3-D} NLTE; thin solid: \mbox{1-D} NLTE, dashed: LTE). The thin vertical line denotes the nominal core wavelength. Note, the 1-D-NLTE profiles coincide almost completely with the 3-D-NLTE profiles.
} 
 \label{fig:avg_profiles}
\end{figure*}
\begin{figure*}
\center{ \resizebox{0.75\hsize}{!}{\includegraphics{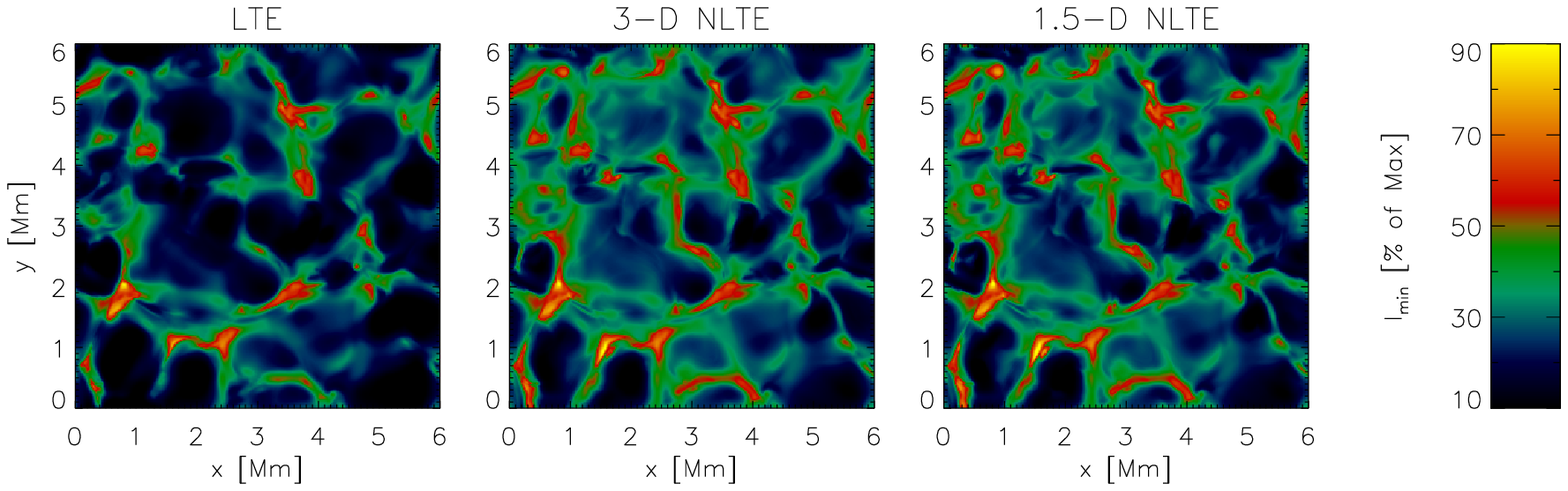}} }
\center{ \resizebox{0.75\hsize}{!}{\includegraphics{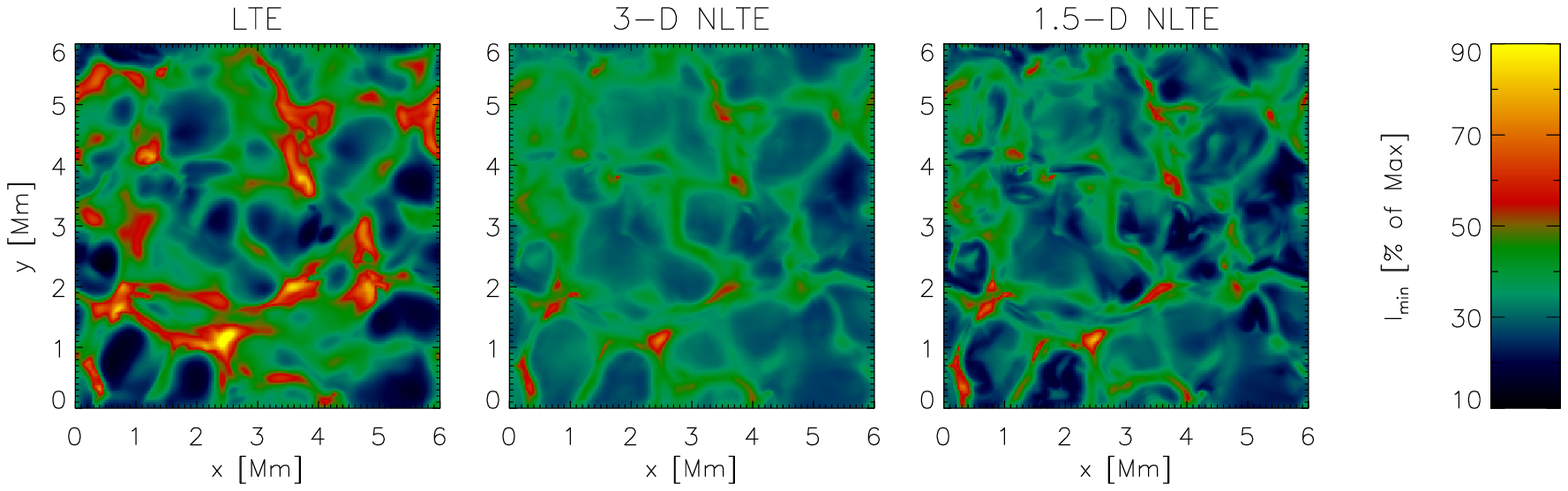}} }
\center{ \resizebox{0.75\hsize}{!}{\includegraphics{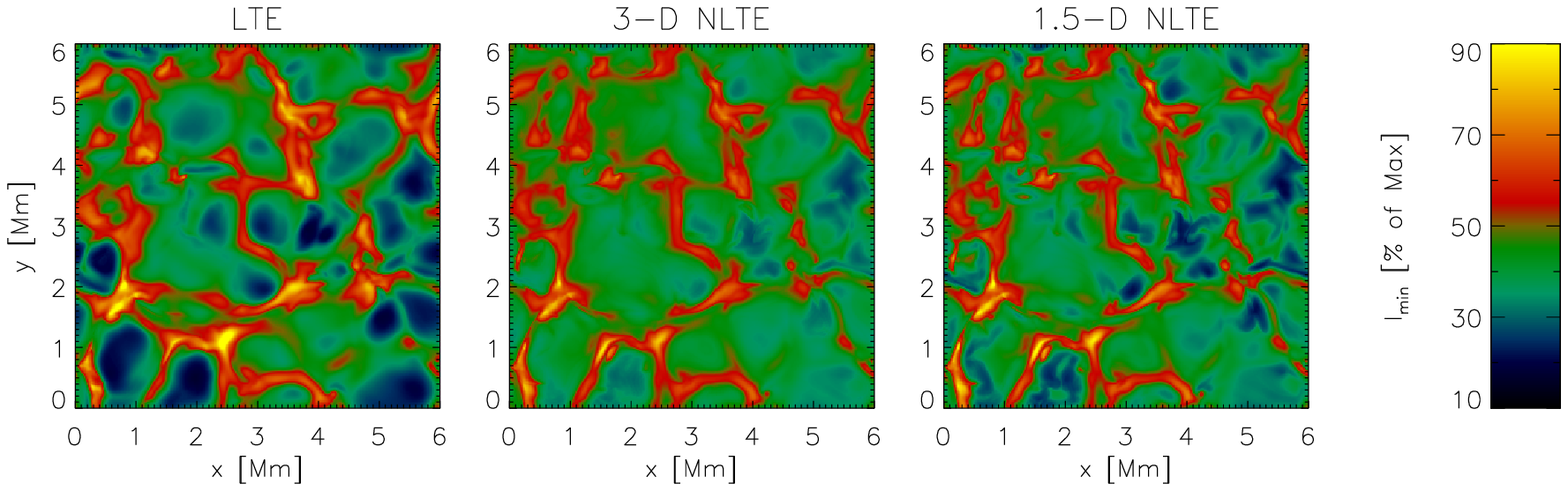}} }
 \caption{ 
 Map of the residual intensity of our three selected lines (top: $525.02$~nm; middle: $630.15$~nm; bottom: $630.25$~nm) for the three different calculation methods (left: LTE; middle: 3-D-NLTE; right: 1-D-NLTE). 
} 
 \label{fig:tv_Imin}
\end{figure*}

\subsection{Computational setup }\label{sec:hdfe23_num_resources}
%%%%%%%%%%%%%%%%%%%%%%%%%%%%%%%%%%%%%%%%%%%%%%%%%%%%%%%%%%%%%%%%%%%%%%%%

For the \mbox{3-D} NLTE calculation, we used the standard angle set A6 of the RH code with 6 rays per octant or 48 in total. This choice is, although not ideal (see Paper~I), reasonable for our case as we only want to compare different calculation methods (LTE, \mbox{3-D} NLTE and \mbox{1-D} NLTE) here. Furthermore, as the original resolution used in the HD simulation was maintained for our calculation, we expect relevant features to be larger than a single pixel and therefore a smaller angular resolution to be acceptable. The choice of 48 angles allows us to reduce the size of the numerical problem considerably. The corresponding \mbox{1-D} runs were done with 5 Gaussian angles.

The size of the numerical problem of a single iteration is given by the size of the atmosphere ($288 \times 288 \times 72$) times the number of rays to be calculated (48) times the number of wavelengths in the spectrum ($\approx 1300$). This circumscribes the number of formal solutions necessary per iteration (roughly $4 \times 10^{11}$). The calculations were executed on a mid-scale server with four 4-threaded CPU's, $128$ GB of memory, and $5$ TB of disk space. The duration of the \mbox{3-D} NLTE run (using 16 threads in parallel) amounted to approximately $1.5$ month. 

For all calculation methods, several rays of different inclination (one with $\mu_z=1.0$) were calculated throughout the cube after convergence. Along these rays, the source function and the optical depth were determined (for a selection of wavelengths only) and they were used to determine contribution functions, emergent profiles and optical path lengths as used in later sections.

%%%%%%%%%%%%%%%%%%%%%%%%%%%%%%%%%%%%%%%%%%%%%%%%%%%%%%%%%%%%%%%%%%%%%%%%
\section{Results}\label{sec:hdfe23_results}
%%%%%%%%%%%%%%%%%%%%%%%%%%%%%%%%%%%%%%%%%%%%%%%%%%%%%%%%%%%%%%%%%%%%%%%%
As in Paper~I, our main focus lies on the differences of the spectral line profiles resulting from the three tested calculation methods (LTE, \mbox{1-D} NLTE, and \mbox{3-D} NLTE) and their origins.

\subsection{Spatial variation of the residual intensity and average profiles}\label{sec:hdfe23_results_average}
%%%%%%%%%%%%%%%%%%%%%%%%%%%%%%%%%%%%%%%%%%%%%%%%%%%%%%%%%%%%%%%%%%%%%%%%
\begin{figure}
\center{ \resizebox{\hsize}{!}{\includegraphics{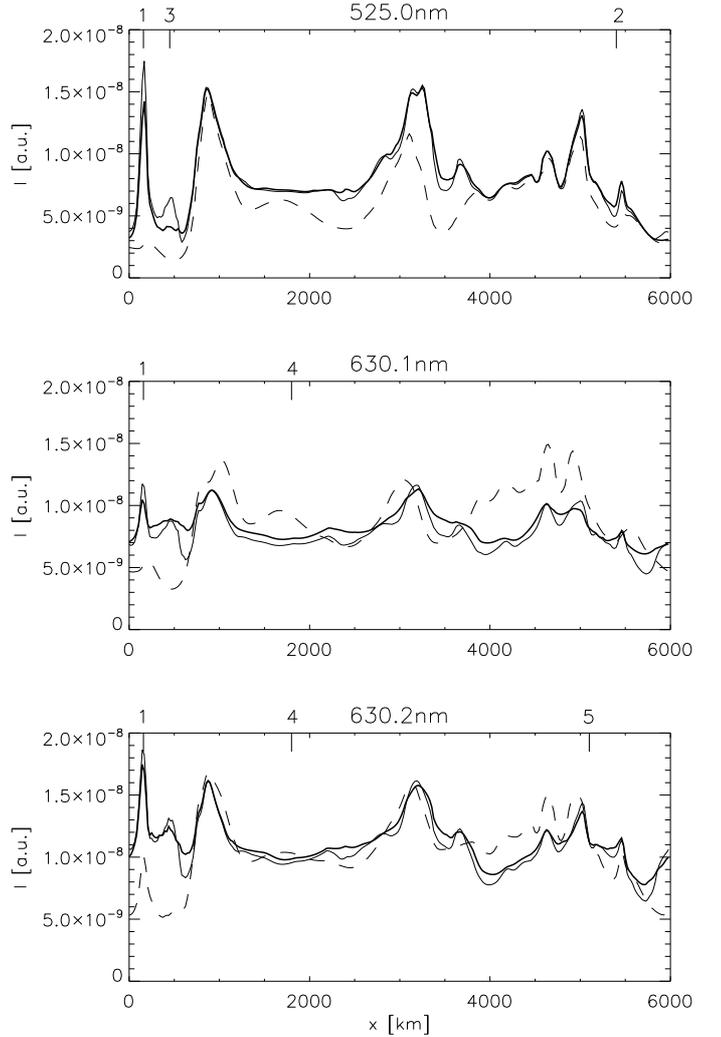}} }
 \caption{ 
Sample cross-sections along the $x$-axis (at $y \approx 2500$~km in Fig.~\ref{fig:tv_atmos}) showing the minimal residual intensity for three lines as indicated above each panel and three methods (thick solid: \mbox{3-D} NLTE; thin solid: \mbox{1-D} NLTE, dashed: LTE). Local intensity maxima correspond to intergranular lanes (see Fig.~\ref{fig:tv_atmos}). The numbered indicators on top of the panels mark $x$ locations discussed in the text.
} 
 \label{fig:sample_profiles}
\end{figure}

The resulting average profiles of our three selected lines are presented in Fig.~\ref{fig:avg_profiles}. The \mbox{1-D} NLTE and the \mbox{3-D} NLTE profiles coincide almost completely for all three lines. The LTE profiles agree fairly well too with the NLTE profiles, seemingly confirming the often-made assumption that these lines are \emph{"built in LTE"}. However, some discrepancy may already be seen in the profile of the $630.25$~nm line and --- more strongly --- in those of the $524.71$/$525.02$~nm line pair. The good agreement between the LTE and NLTE profiles is further reduced if one considers the (resolved) line core residual intensity maps, which are given in Fig.~\ref{fig:tv_Imin}. Note that the color scales used for a single line are the same for all calculation methods. Differences between the images may therefore be directly attributed to the calculation method. 

The spatial variation of the residual intensity in the three lines shows considerable differences between the three methods. Thus the contrast is strongest in LTE and weakest in \mbox{3-D} NLTE in all three lines shown, with the difference between \mbox{1-D} NLTE and \mbox{3-D} being smaller than that between the LTE and NLTE calculations. This corroborates our findings of Paper~I. However, as we do not consider a magnetic atmosphere, we expect that the effects tend to be smaller than in our flux tube/sheet model of Paper~I. Also, the finite horizontal resolution of the HD cube should lead to smaller differences than in Paper~I.

To provide a more quantitative view of the variability of the line profiles calculated in LTE, \mbox{1-D} NLTE and \mbox{3-D} NLTE we present, in Fig.\ref{fig:sample_profiles}, the residual intensity along a cut through the atmosphere (at $y \approx 2500$~km). Some $x$ locations on which we like to draw attention are marked with a numbered indicator on top of the individual panels.

The LTE residual intensity in the $525$~nm lines (top panel) often --- but not always --- lies clearly below the NLTE intensities. For this line pair, the intensity difference between \mbox{1-D} and \mbox{3-D} NLTE RT is small. Significant differences are only found at a few spatially well-isolated intensity extrema (typically located near or in intergranular lanes, such as the first maximum near $x=150$~km (Pos. 1), see also Fig.\ref{fig:tv_atmos}). There, the variation of the \mbox{1-D} NLTE intensity is often --- but not always --- larger. This corresponds to our findings in Paper~I and those of \citet{stenholmstenflo1977}. The horizontal radiation in the \mbox{3-D} calculation weakens lines at locations where the intensity of the environment is stronger (i.e. at local minima of $I$ as e.g. at Pos. 2 near $x \approx 5400$~km). At local intensity maxima, the environment is darker (cooler) and correspondingly, the inverse effect, i.e. a reduction of the \mbox{3-D} intensity, takes place (Pos. 1). As the \mbox{1-D} calculation neglects both types of contributions of the horizontal RT, we expect the contrast to be larger in \mbox{1-D} NLTE than in \mbox{3-D} NLTE. This can be confirmed from Fig.~\ref{fig:tv_Imin}. 

Note that there are locations in Fig.~\ref{fig:sample_profiles} where the above statement is obviously not valid (Pos. 3). This may be explained by the fact, that our cut along $x$ in Fig.~\ref{fig:sample_profiles} shows only the neighborhood (in intensity) in one direction and we therefore miss part of the necessary information on the environment of the pixels in question. Thus, some of the local maxima seen in Fig.~\ref{fig:sample_profiles} are no longer maxima when considered in another direction.

In the $630$~nm lines, the effect of horizontal RT is larger. The locations where the \mbox{1-D} NLTE and \mbox{3-D} NLTE differ are more numerous. Furthermore and in contrast to the lines near $520$~nm, the LTE intensity can be higher or lower than the corresponding NLTE intensities. The intensity variation (contrast) is again largest in LTE, and that of \mbox{1-D} NLTE is larger than that of \mbox{3-D} NLTE, as expected. If one compares the $630.15$~nm in Fig.~\ref{fig:sample_profiles} with the temperature distribution at $y=2500$~km and $z=240$~km in the right panel of Fig.~\ref{fig:tv_atmos}, one may discover a tendency for the \mbox{1-D} NLTE intensity to be lower than the \mbox{3-D} NLTE intensity in the (colder) granular areas (Pos. 4) and to be higher than the \mbox{3-D} NLTE intensity in the (hotter) intergranular lanes (Pos. 1 \& 5). Thus, the \mbox{3-D} line is weakened in the granules by the strong irradiation from its environment and strengthened in the intergranular lanes by the weaker irradiation from the colder granules. Because the lines are formed above the granular temperature inversion, the effect is reversed if compared to the case of a flux tube or sheet. However, the diversity of possible situations is large. Many locations exist where the three profiles completely coincide, while at other positions clear differences with either sign exist. The main reason for such exceptions is the fact that irradiation from other heights (not visible in Fig.~\ref{fig:tv_atmos}) may influence the line formation as well (vertical gradients of the temperature are not always well-correlated with the temperature at a given height, e.g. at $z=240$~km).

\subsection{Contrasts and their center-to-limb variation}\label{sec:hdfe23_result_contrast}
%%%%%%%%%%%%%%%%%%%%%%%%%%%%%%%%%%%%%%%%%%%%%%%%%%%%%%%%%%%%%%%%%%%%%%%%
\begin{figure*}
\center{ \resizebox{0.75\hsize}{!}{\includegraphics{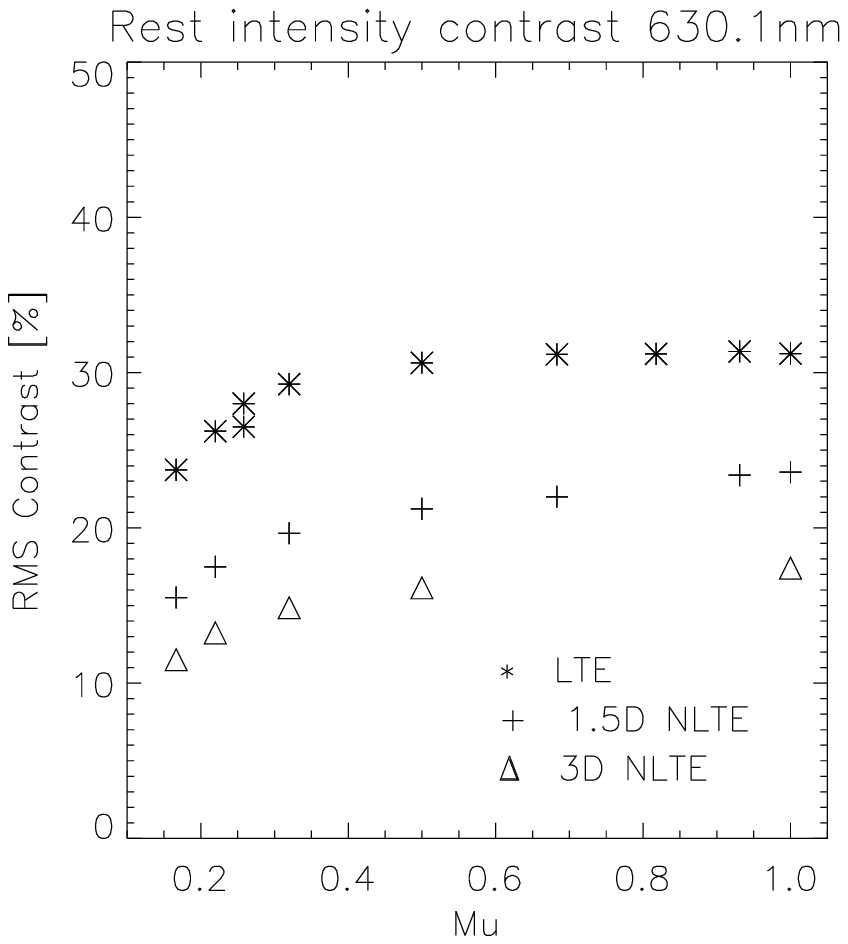}
                                   \includegraphics{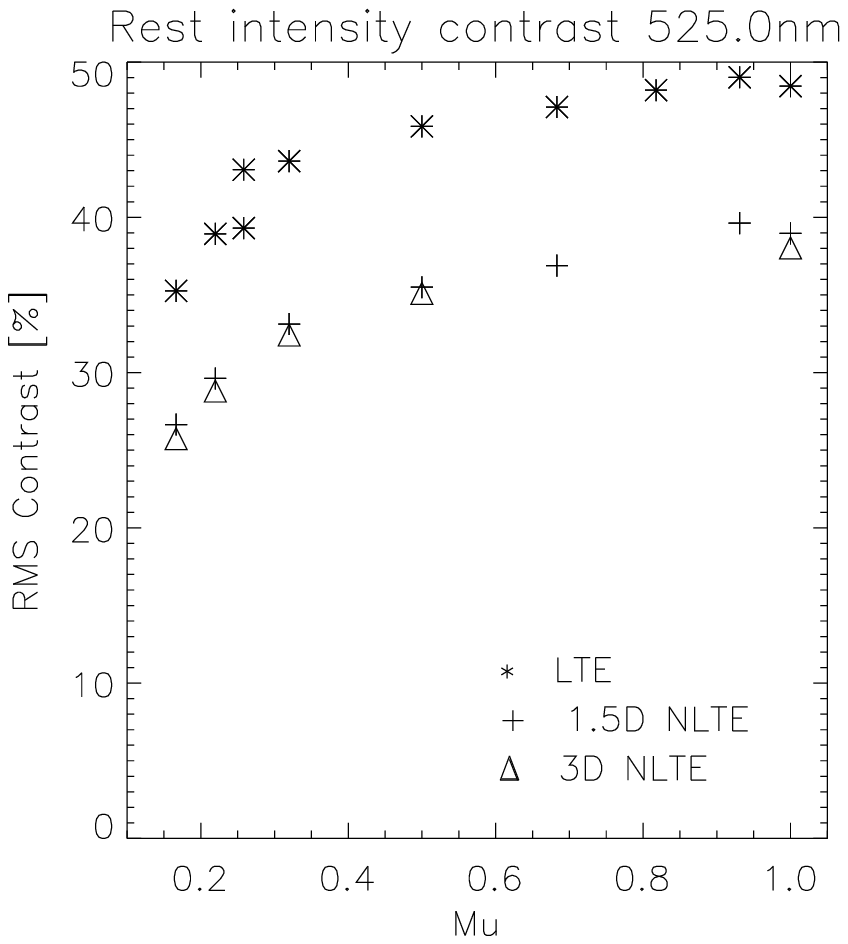} }}
 \caption{Center-to-limb variation of the minimum residual intensity contrast of the $630.15$~nm (left panel) and the $525.02$~nm (right panel) lines for the three calculation methods (stars: LTE; triangles: 3-D-NLTE; plus signs: 1-D-NLTE). 
} 
 \label{fig:clv_Imin}
\end{figure*}

\begin{figure*}
\center{ \resizebox{0.75\hsize}{!}{\includegraphics{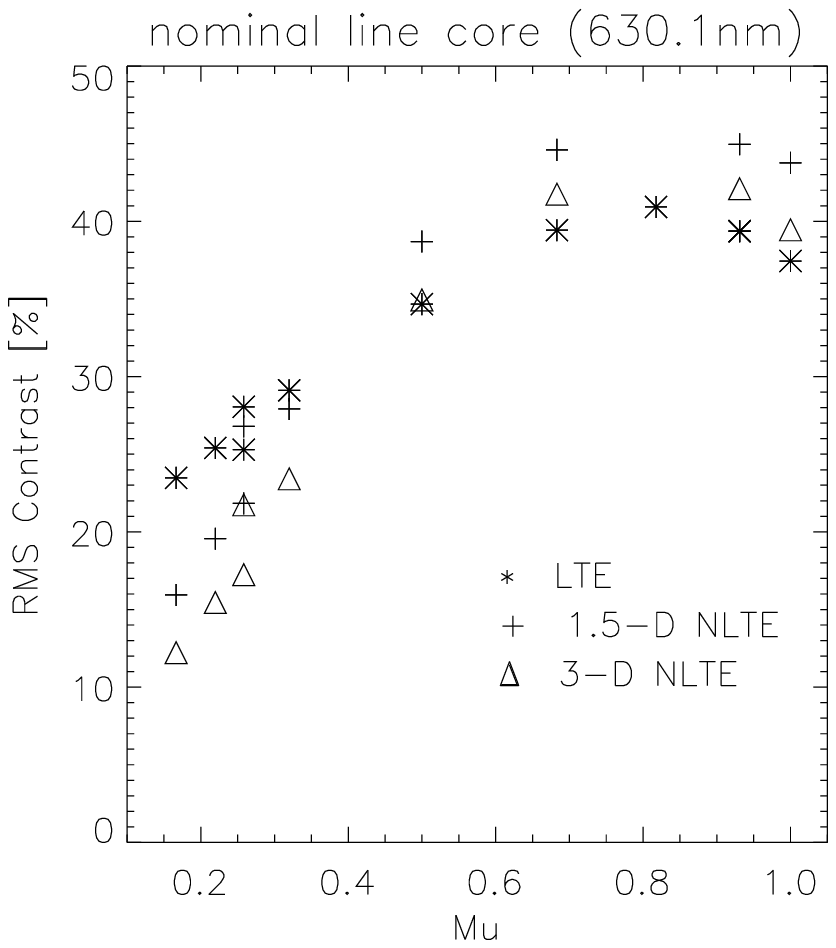}
                                   \includegraphics{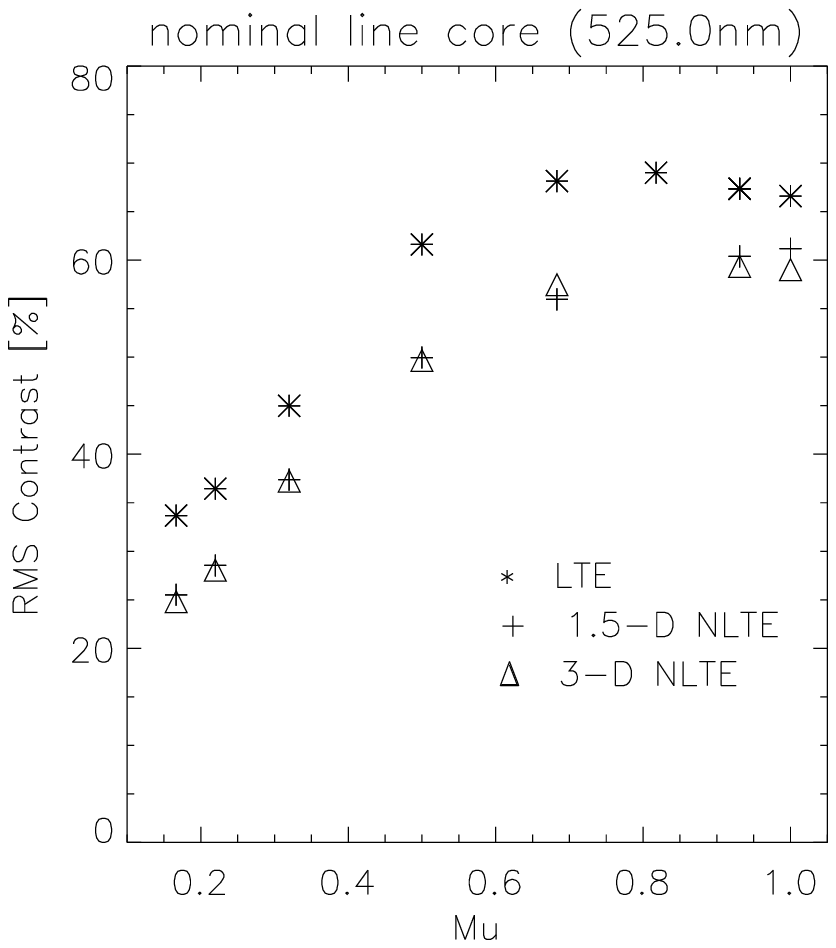} }}
 \caption{The same as Fig.~\ref{fig:clv_Imin}, but for the contrast at the nominal core wavelength. Note the different contrast scales of the left and right panel. 
} 
 \label{fig:clv_I0}
\end{figure*}

In the previous section, we found qualitative evidence for a strong dependence of the contrast at $\mu=1.0$ on the calculation method. In this section we quantify these differences for various inclinations of the line of sight (LOS) to the solar surface normal. 

We begin by considering the (normalized) RMS of the residual intensity
\begin{equation}\label{eq:contrast}
I_{RMS}
\,\,=\,\,
\frac{1}{\langle I_{min} \rangle} \sqrt{ \frac{ 1 }{N_{xy}} \sum \limits_{(x,y)} (I_{min}(x,y) - \langle I_{min} \rangle)^2 } 
\rm \ ,
\end{equation}
with $N_{xy}$ the number of grid points in the $(x,y)$ plane, $I_{min}$ the residual intensity (i.e. the minimum intensity in the line), and $\langle I_{min} \rangle$ its spatial average. Due to Doppler shifts this intensity refers to different wavelengths at different spatial locations. Already Fig.~\ref{fig:tv_Imin} suggested that the RMS contrast is largest in LTE. This is confirmed by Fig.~\ref{fig:clv_Imin}, where the CLV of the RMS contrast of residual intensity in the $630.15$~nm (left panel) and the $525.02$~nm (right panel) lines are plotted. 

The contrasts in Fig.~\ref{fig:clv_Imin} increase with $\mu$ only until $\mu=0.3$. For larger $\mu$ they are almost independent of $\mu$ in case of the $630.15$~nm line or only slightly increase with $\mu$ ($525.02$~nm line). The LTE contrasts are larger than the \mbox{1-D} NLTE contrasts at all $\mu$, which are for \ion{Fe}{i} $630.15$~nm in turn much larger than those obtained with \mbox{3-D} RT. For the $525.02$~nm line the \mbox{1-D} and \mbox{3-D} contrast are virtually the same for all LOS. The differences for the $630.15$~nm line are surprisingly large: The \mbox{1-D} NLTE contrasts are 30\% ($\mu >= 0.3$) to 50\% (small $\mu$) larger than those of the \mbox{3-D} RT. The contrasts computed in LTE are approximately a factor of two larger than those obtained with \mbox{3-D} NLTE, independent of $\mu$. The absolute values of $I_{RMS}$ of the $525.02$~nm line are significantly higher that of $630.15$~nm line. $I_{RMS}$ for the $630.25$~nm line (not plotted) agree qualitatively well with those of the $630.15$~nm line, although the differences between the three calculation methods are slightly smaller ($\approx 20$\% for the difference between \mbox{1-D} and 3-D, $50$ - $70$\% for those between LTE and 3-D). The larger $I_{RMS}$ of the $525$~nm line pair is due to the large temperature sensitivity of these lines stemming from their lower excitation potential. The absolute difference between the methods and between the lines are only very weakly dependent of $\mu$.

We now attempt to explain the general shape of the center-to-limb variation (CLV). The plateau in the residual intensity contrast at Fig.~\ref{fig:clv_Imin} at values of $\mu$ above $0.3$ and the drop in $I_{RMS}$ for smaller $\mu$ may be attributed to the following facts. The variation of $I_{min}$ originates mainly from horizontal temperature variations in the layers in which $I_{min}$ is formed. The height of this region is only weakly dependent on the LOS angle, especially for larger $\mu$. In addition, the temperature variation decreases only slowly with height at these heights, leading to the plateau. The drop in $I_{RMS}$ at $\mu < 0.3$ is mainly caused by a geometrical effect: The structuring of the photosphere into granules and intergranular lanes with their strong temperature differences can be seen only at large $\mu$. For small $\mu$, we do not see this structuring anymore. The $\tau=1$ surface at the wavelength of the intensity minimum is then much more homogeneous and so is the temperature distribution. 

A considerably different behavior is observed if instead of $I_{min}$ we consider the intensity at the fixed wavelength $\lambda_0$ corresponding to the nominal line core, i.e. the core wavelength of the spatially averaged profile. This case is of interest since it mimics a fixed-$\lambda$ narrow-band filter. Figure~\ref{fig:clv_I0} depicts the CLV of the intensity contrasts at the nominal line-core wavelength ($\lambda_0$) of the same lines as in Fig.~\ref{fig:clv_Imin}. These contrasts often (but not always) peak around $\mu \approx 0.7$, dropping slowly towards disc center and rather rapidly towards the limb for all methods. As in the case of $I_{min}$ and for mainly the same reason, the $525.02$~nm line contrast is higher than that of the $630.15$~nm line, reaching values as high as $70$\% (note the different scales in the two panels of Fig.~\ref{fig:clv_I0}).

As in Fig.~\ref{fig:clv_Imin}, the \mbox{1-D} NLTE contrasts for $630.15$~nm in Fig.~\ref{fig:clv_I0} (plus signs) are significantly higher than those of the \mbox{3-D} calculation (triangles) for all $\mu$, whereas those of the $525.02$~nm line almost coincide with each other. In Fig.~\ref{fig:clv_I0}, the LTE contrast of $630.15$~nm drops off less rapidly towards lower $\mu$ than the NLTE contrasts. For $\mu \ge 0.5$ the LTE contrast is actually lower than the NLTE contrasts, but much larger for small $\mu$. The same behavior, although slightly weaker, is exhibited by the $630.25$~nm line (not shown here), whereas for the $525$~nm lines the LTE and NLTE contrasts run more in parallel.

\begin{figure}
\center{ \resizebox{\hsize}{!}{\includegraphics{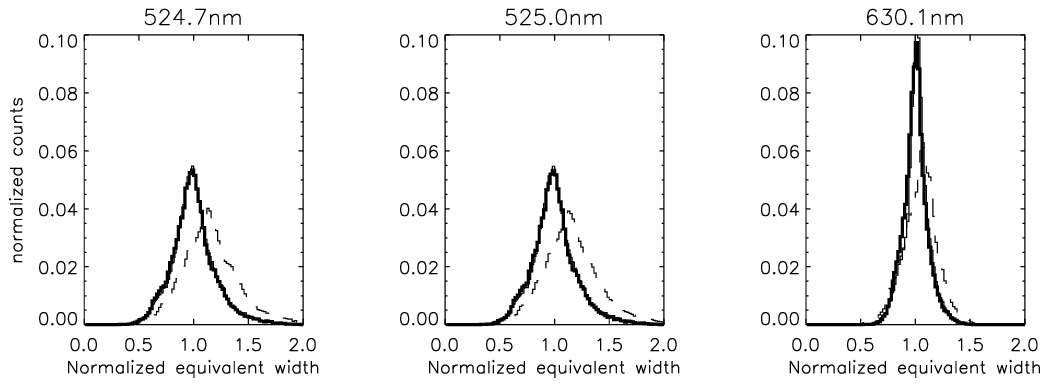}} }
 \caption{ 
Distribution of the equivalent widths normalized to the average \mbox{3-D} NLTE equivalent width. Line styles for the three methods as in Fig.~\ref{fig:sample_profiles}. 
} 
 \label{fig:histo_Aeqw}
\end{figure}

The difference to Fig.~\ref{fig:clv_Imin} comes from the fact that the contrasts at $\lambda_0$ for large $\mu$ are strongly affected by Doppler shifts due to the strong up- and down-flows, which cause the line core to shift away from its nominal position. The nominal line core wavelength may then lay closer to the steep line flanks and already a small wavelength shift may result in a large intensity increase. The magnitude of this effect depends on how narrow the line core is. The LTE line is slightly broader (see Fig.~\ref{fig:histo_Aeqw}) and it is formed slightly higher than the NLTE line. With increasing height, the decreasing RMS variation of the vertical velocities further reduces the LTE contrast. For small $\mu$, however, the line is mainly sensitive to the temperature distribution at the height of formation, because the line is generally broader with less steep flanks and therefore less sensitive to the --- now horizontal and more homogeneous --- velocity field. Hence, the contrasts for low $\mu$ are the same in both, Figs.~\ref{fig:clv_Imin} and \ref{fig:clv_I0}. The difference between the three calculation methods is then mainly caused by NLTE effects and horizontal RT, both of which lower the contrast. 

The reason for the strong differences between contrasts resulting from the \mbox{1-D} and the \mbox{3-D} calculation even at low $\mu$ (in both, Figs.~\ref{fig:clv_Imin} and \ref{fig:clv_I0} may be explained by the fact that in 3-D, although the observer cannot see the horizontal structuring of the granules anymore, the individual atoms emitting the observed photons still see their own environment, including the horizontal variation of the temperature and of the incident radiation. As the horizontal RT typically lowers the contrast in the line by changing the population numbers of the relevant atomic levels, the resulting emitted intensity is accordingly changed independently of the direction of the emitted photon.

The main result of this section is that the granular RMS contrasts in \ion{Fe}{i} line cores may be very large and are strongly affected by \mbox{3-D} NLTE effects.
% -- ALTE VERSION OHNE CLV --

%From Fig.~\ref{tv_Imin } but also Fig.~\ref{fig:sample_profiles} it is evident, that the contrasts calculated with the three methods must differ significantly. Table \ref{tab:contrasts} gives an overview over the RMS contrasts in the three selected lines for the three methods at two different wavelengths, namely the nominal line core and at the effective line core. Although the difference between \mbox{1-D} und \mbox{3-D} NLTE is generally smaller than that between LTE and \mbox{3-D} NLTE, it may e.g. for the $630.1$~nm line amount to remarkable 35\%. Note that the \mbox{1-D} contrasts are always larger than those of \mbox{3-D} NLTE, as expected from horizontal RT which typically levels out contrasts.

%\begin{table}
%\caption{RMS contrast in three selected lines for three different methods. Upper part: RMS contrast in the %minimum residual intensity; lower part: RMS contrast at nominal line core wavelength.}
%\begin{tabular}{lrrrr}
%\hline
% Line & \mbox{3-D} NLTE & LTE & \mbox{1-D} NLTE \\
%\hline
%$525.0$~nm & 0.38 & 0.48 & 0.39 \\
%$630.1$~nm & 0.17 & 0.31 & 0.24 \\
%$630.2$~nm & 0.19 & 0.29 & 0.22 \\
%\hline
%$525.0$~nm & 0.59 & 0.67 & 0.61 \\
%$630.1$~nm & 0.39 & 0.36 & 0.44 \\
%$630.2$~nm & 0.43 & 0.40 & 0.45 \\
%\hline
%\end{tabular}
%\label{tab:contrasts}
%\end{table}

%5250 & 0.37976867 & 0.48443276 & 0.38958978 \\
%6301 & 0.17299066 & 0.31212463 & 0.23591871 \\
%6302 & 0.18589849 & 0.28856537 & 0.22209945 \\
%\hline
%5250 & 0.58924083 & 0.66750495 & 0.61157601 \\
%6301 & 0.39309012 & 0.36418131 & 0.43735293 \\
%6302 & 0.42464749 & 0.39635855 & 0.45283219 \\

\subsection{Equivalent widths, line depths and wavelength shifts}\label{sec:hdfe23_results_other_quantities}
%%%%%%%%%%%%%%%%%%%%%%%%%%%%%%%%%%%%%%%%%%%%%%%%%%%%%%%%%%%%%%%%%%%%%%%%

\begin{figure}
\center{ \resizebox{0.85\hsize}{!}{\includegraphics{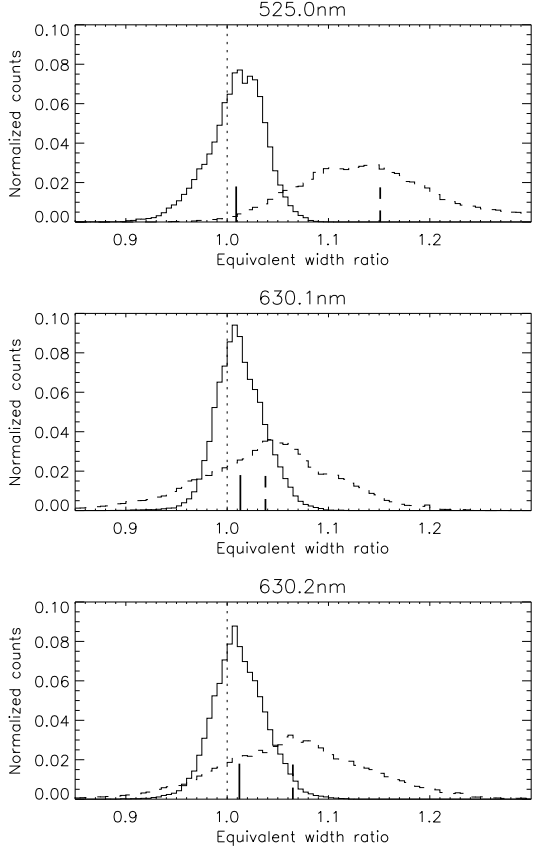}} }
 \caption{ 
Distribution of the equivalent width ratios (solid: $W_{1-D NLTE}/W_{ \mbox{3-D} NLTE}$; dashed: $W_{LTE}/W_{ \mbox{3-D} NLTE}$) for three lines as indicated on top of the panels. The short, thick vertical lines denote the average value of the two ratios.
}
 \label{fig:histo_ratio_Aeqw}
\end{figure}

\begin{figure*}
\center{ \resizebox{0.7\hsize}{!}{\includegraphics{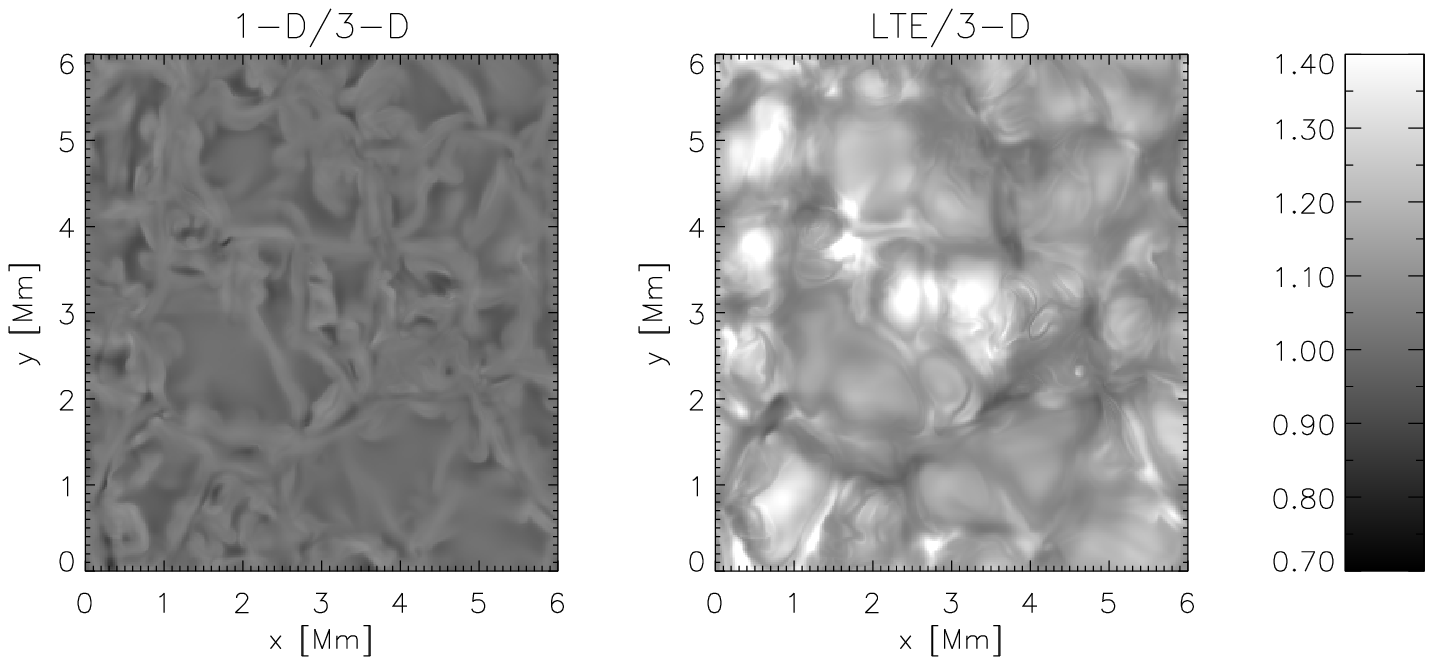}} }
\center{ \resizebox{0.7\hsize}{!}{\includegraphics{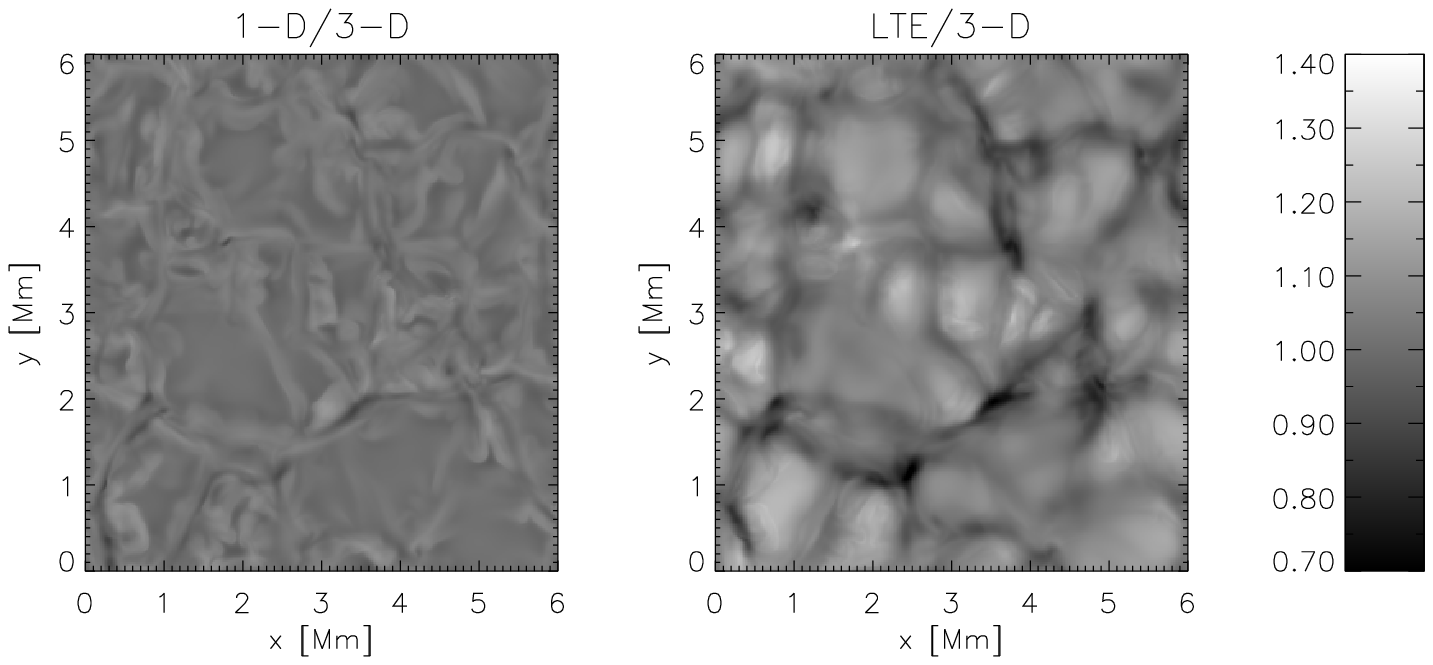}} }
\center{ \resizebox{0.7\hsize}{!}{\includegraphics{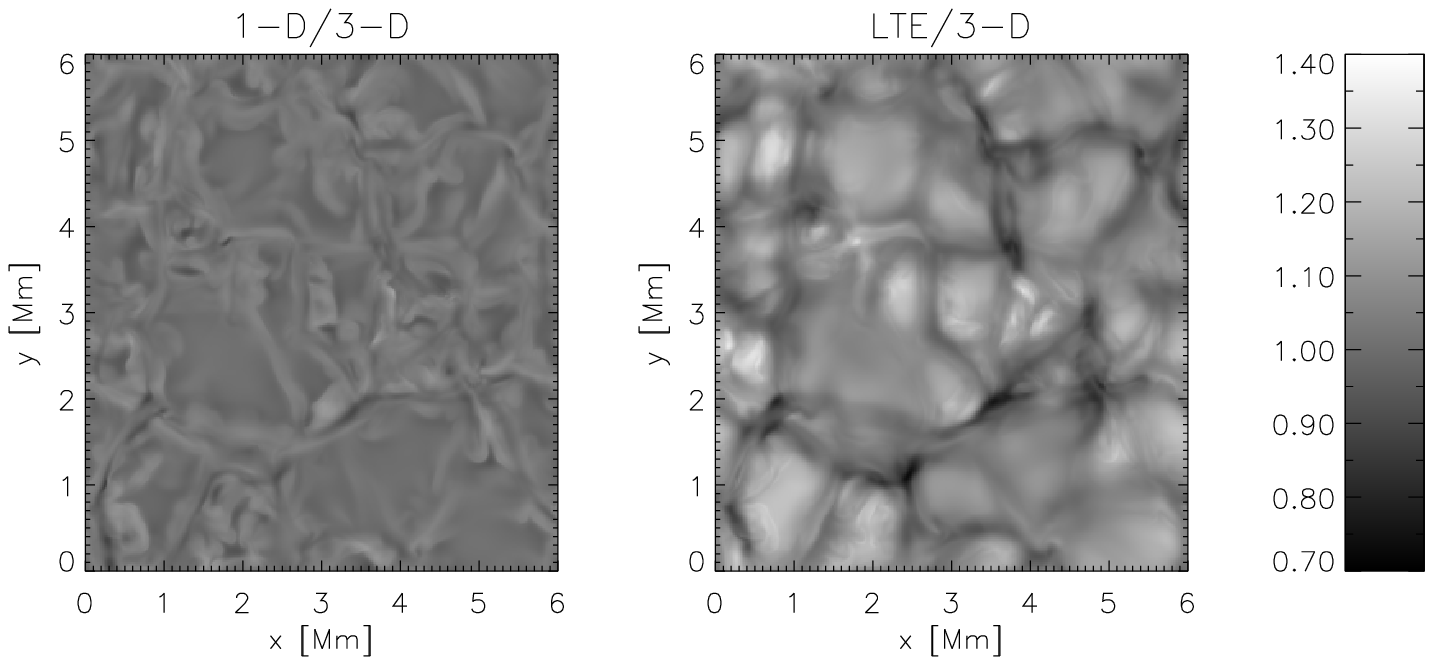}} }
 \caption{ 
 Map of the equivalent widths relative to the 3-D-NLTE value for the same lines as in previous figures. Left: 1-D-NLTE, Right: LTE.  
} 
 \label{fig:tv_ratio_Aeqw}
\end{figure*}

\subsubsection{Equivalent width}\label{sec:hdfe23_results_other_quantities_EW}
%%%%%%%%%%%%%%%%%%%%%%%%%%%%%%%%%%%%%%%%%%%%%%%%%%%%%%%%%%%%%%%%%%%%%%%%
% durchschnittswert: 5250 hat langen schwanz da teilweise in LTE wohl fast in Emission und dann das Ratio schnell sehr gross, ie wenn \mbox{3-D} Linie sehr schwach wird das Ratio sehr schnell riesig.
We begin this section with equivalent widths, $W$, due to their importance for the computation of elemental abundances. Figure~\ref{fig:histo_Aeqw} displays the distribution of equivalent widths for the same lines and calculation methods as in Figs.~\ref{fig:tv_Imin} and \ref{fig:sample_profiles}. As was the case for the spatially averaged line profiles, the distributions look --- at a first glance --- quite similar, although some differences between LTE and NLTE distributions are seen. However, the similarity of the distributions can mask the fact that at any given pixel $W$ can be rather different. To quantify these differences on a pixel by pixel basis, we calculate the ratios $W_{LTE}/W_{ \mbox{3-D} NLTE}$ and $W_{1-D NLTE}/W_{ \mbox{3-D} NLTE}$ for each pixel in the whole $xy$ plane. The histograms in Fig.~\ref{fig:histo_ratio_Aeqw} present the distribution of these ratios. They manifest that LTE overestimates $W$ by roughly $10$\% on average. \mbox{1-D} NLTE also appears to overestimate $W$, although by a much reduced amount. However, there is a considerable spread in the ratios. The standard deviations of the plotted distributions are $\approx 4$\% for LTE/ \mbox{3-D} and $\approx 2$\% for 1-D/3-D. In extreme cases the $W$ can differ by $10$\% in \mbox{1-D} and 3-D.

The spatial distribution of the equivalent width ratios of Fig.~\ref{fig:histo_ratio_Aeqw} is presented in Fig.~\ref{fig:tv_ratio_Aeqw}. The largest disagreement is --- not unexpectedly --- found at the edge of the granules, where large temperature and intensity gradients are found.

\subsubsection{Line depth}\label{sec:hdfe23_results_other_quantities_linedepth}
%%%%%%%%%%%%%%%%%%%%%%%%%%%%%%%%%%%%%%%%%%%%%%%%%%%%%%%%%%%%%%%%%%%%%%%%
In complete analogy to the equivalent width, differences in the line depth calculated with the three methods are of the same magnitude, i.e. typically 10\% - 20\% between LTE and \mbox{3-D} NLTE (not shown here). The largest differences are again found near intergranular lanes. 

\begin{figure}
\center{ \resizebox{0.85\hsize}{!}{\includegraphics{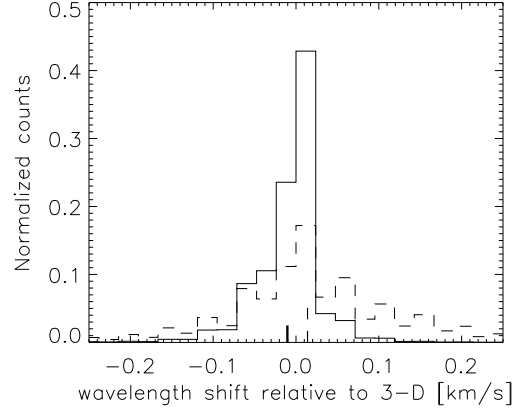}} }
 \caption{ 
Distribution of the core wavelength $\lambda_{min}$ shifts relative to the \mbox{3-D} NLTE value ($630.25$~nm line). Solid line: \mbox{1-D} -- \mbox{3-D} ; dashed line: LTE -- 3-D. Averages are indicated by the small vertical lines (as in Fig.~\ref{fig:histo_ratio_Aeqw}).
} 
 \label{fig:wavelenshift}
\end{figure}

\subsubsection{Doppler shift}\label{sec:hdfe23_results_other_quantities_doppler}
%%%%%%%%%%%%%%%%%%%%%%%%%%%%%%%%%%%%%%%%%%%%%%%%%%%%%%%%%%%%%%%%%%%%%%%%
The statistical distribution of the Doppler shifts (determined by the difference between the average wavelength position of the two line flanks at half height and the nominal line core) relative to the \mbox{3-D} NLTE value for the $630.25$~nm line are depicted in Fig.~\ref{fig:wavelenshift}. The Doppler shift distributions in the other lines are of the same size or even smaller and therefore not shown here. Almost no $(x,y)$ position shows any significant difference in the Doppler shift between any two methods. More than $80$\% of the \mbox{1-D} NLTE pixel have a Doppler shift which does not differ by more than $100$ m/s from the \mbox{3-D} NLTE value. The distribution of the relative differences of the LTE Doppler shifts is slightly broader, which may be explained by the slightly higher opacity of the LTE line, so that it samples a somewhat higher region in the atmosphere (see e.g. Fig.\ref{fig:cut_dT_contrib}) and may therefore measure slightly different velocities.

\subsection{Origin of the differences between \mbox{1-D} and \mbox{3-D} NLTE radiative transfer}\label{sec:hdfe23_results_origin}
%%%%%%%%%%%%%%%%%%%%%%%%%%%%%%%%%%%%%%%%%%%%%%%%%%%%%%%%%%%%%%%%%%%%%%%%
\begin{figure*}
\center{ \resizebox{0.85\hsize}{!}{\includegraphics{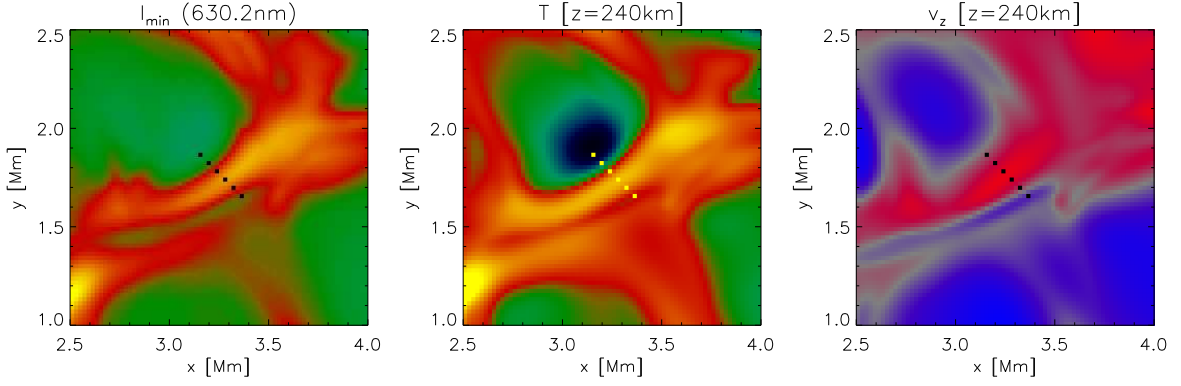}} }
 \caption{ 
Enlarged sub-section of the atmosphere. The color scales ($T$, $v_z$) as well as the $x$ and $y$ axis correspond to those in Fig.~\ref{fig:tv_atmos} and \ref{fig:tv_Imin} (residual intensity of the $630.25$~nm line) respectively. The black and yellow dots indicate the positions selected in this section for a more detailed analysis.
} 
 \label{fig:tv_atmos_blowup}
\end{figure*}

\begin{figure*}
\center{ \resizebox{0.8\hsize}{!}{\includegraphics{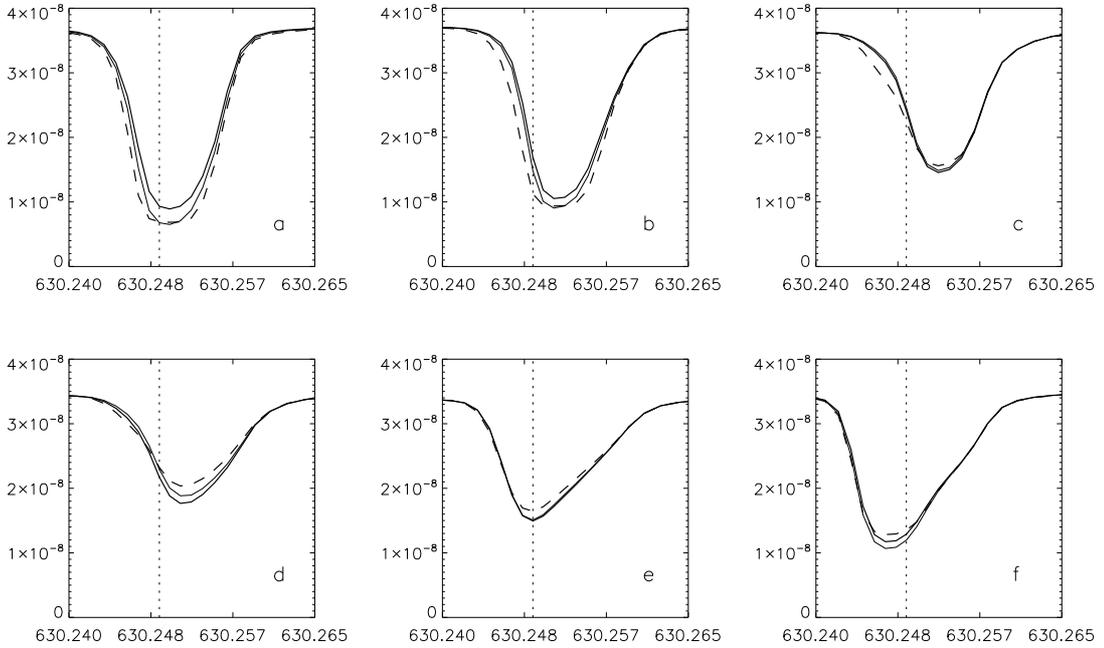}} }
 \caption{ 
Line profiles of the $630.25$ line at six different spatial positions along a cut through our model (see Fig.~\ref{fig:tv_atmos_blowup}, where the cut is shown. The panels from left to right and top to bottom correspond to positions a to f along the cut, i.e. from its top-left to its bottom-right corner. Thick solid: \mbox{3-D} NLTE; thin solid: \mbox{1-D} NLTE, dashed: LTE.
} 
 \label{fig:cut_profiles}
\end{figure*}

\begin{figure*}
\center{ \resizebox{0.8\hsize}{!}{\includegraphics{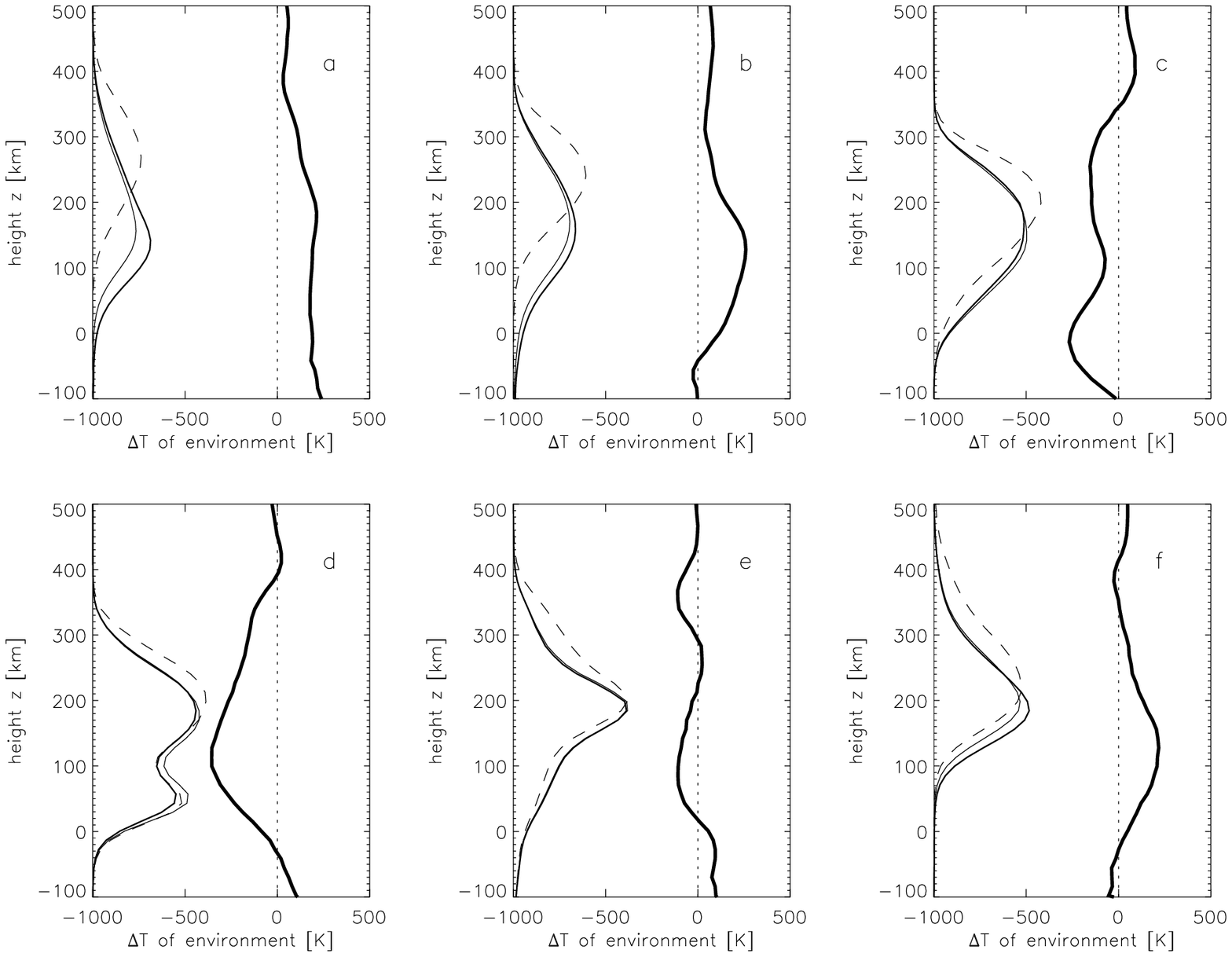}} }
 \caption{ 
Thick line at right side of each panel: Temperature difference, $\Delta T$, of the line-forming region relative to its environment for the same positions as in Fig.~\ref{fig:cut_profiles}. See main text for the exact definition of $\Delta T$. Positive $\Delta T$: the environment is hotter than the line-forming region. 
Lines on the left side of each panel: Normalized contribution functions for the three methods indicating the heights of line formation (thick solid: \mbox{3-D} NLTE; thin solid: \mbox{1-D} NLTE, dashed: LTE).
} 
 \label{fig:cut_dT_contrib}
\end{figure*}

In this section we will show that the differences between \mbox{1-D} NLTE and \mbox{3-D} NLTE have their origin in the effect first described by \citet{stenholmstenflo1977}, i.e. in the horizontal irradiation from hotter (or cooler) environments into the line-forming region. To see this we examine one small region of our atmosphere in more detail. We selected a set of 6 points along a cut through an intergranular lane as indicated in Fig.~\ref{fig:tv_atmos_blowup}, the latter reproducing a section of $1500 \times 1500$~km$^2$ of the original atmosphere. The units of the $x$ and $y$ axes in Fig.~\ref{fig:tv_atmos_blowup} are the same as those in Fig.~\ref{fig:tv_atmos}. Every second pixel was chosen along the diagonal cut which starts at the upper left (in the following called position a) and ends at the lower right (position f in the following). The spatial distance between two such pixels amounts to approximately $55$~km.

Figure~\ref{fig:cut_profiles} shows the profiles along the cut. The panels exhibit \mbox{1-D} NLTE (and LTE) profiles that differ clearly from the \mbox{3-D} NLTE profiles, as well as from each other. Also, the sign of the difference changes along the cut. Whereas the \mbox{3-D} solution gives the weakest spectral line in panels $a$ and $b$, it results in the strongest line in panel d. In the other panels the results are not entirely clear-cut. We see also that LTE profiles often lie closer to the \mbox{1-D} NLTE profiles, but not at every location (see, e.g., panel f for an exception).

To understand the circumstances of the line formation in more detail, we have to consider the conditions at heights where the line forms. We define the line-forming region $F_{line}(x,y)$ at the location $(x,y)$ as the heights $z$ where the line contribution function $C_{line}(z)$ is larger than 20\% of its maximum $C_{line}^{max}$, i.e.

\begin{equation}\label{eq:lineformingregion}
F_{line}
\,\,=\,\,
\{z \mid C_{line}(z) > 0.2\times C_{line}^{max} \} 
\rm \ .
\end{equation}

We next define a temperature estimate of the environment, $T^{Env}$, whose difference to the temperature at the spatial pixel in question presumably determines magnitude and sign of the \mbox{3-D} RT effects. As the typical horizontal mean free path of a photon at the average heights of line core formation in the photosphere is about $100$~km, we define as the environment of a point $P_{x_0,y_0,z_0}$ 

\begin{equation}\label{eq:envP}
Env(P_{x_0,y_0,z_0})
\,\,=\,\,
\{P_{x_0\pm\Delta x,y_0\pm\Delta y,z_0}\} 
\rm \ ,
\end{equation}
with $\Delta x=\Delta y=105$~km, which corresponds to a distance of 5 pixel. As the temperature $T^{Env}$ of the environment at a point $P_{x_0,y_0,z_0}$, we define the average temperature of the four pixels in $Env(P_{x_0,y_0,z_0})$, i.e. 

\begin{equation}\label{eq:Tenv}
 T^{Env}(x_0,y_0,z_0) = \sum \limits_{(x,y)=(x_0\pm\Delta x,y_0\pm\Delta y)}T(x,y,z_0) /4 
\rm \ .
\end{equation}

The temperature difference of the environment at a point $P_{x_0,y_0,z_0}$ in the atmosphere simply becomes
\begin{eqnarray}\label{eq:deltaTenv}
\Delta T^{Env}(x_0,y_0,z_0)
\,\,=\,\,
T^{Env}(x_0,y_0,z_0) - T(x_0,y_0,z_0) \\
\rm \ .
\end{eqnarray}

Figure~\ref{fig:cut_dT_contrib} presents the run of $\Delta T^{Env}$ (thick solid line on the right side of each frame) as a function of $z$ for the same six locations as in Fig.~\ref{fig:cut_profiles}. On the left side of each panel, the corresponding contribution functions of $630.25$~nm (the same line as in Fig.~\ref{fig:cut_profiles}) are given for the three methods (thick solid: \mbox{3-D} NLTE, \mbox{1-D} NLTE thin solid, dashed: LTE, for reference only). 

One easily recognizes that temperature differences of several $100$ K may exist between the line-forming region and its environment. At points a, b and f, the environment is significantly hotter, at point d significantly cooler, and, in panels c und e, of similar temperature. If we compare the temperature differences in Fig.~\ref{fig:cut_dT_contrib} with the line profiles in Fig.~\ref{fig:cut_profiles}, we immediately recognize that the \mbox{1-D} NLTE line profiles in panels a, b and f are deeper than those calculated in \mbox{3-D} NLTE. In panel d however, the \mbox{3-D} NLTE line is clearly stronger, and in panels c and e the two profiles show only minor differences. This is exactly what one would expect from the influence of horizontal RT. Radiation from relatively hot environments weakens the \mbox{3-D} NLTE line profiles and colder environments may lead to the inverse effect, i.e. a strengthening of the \mbox{3-D} NLTE line, as in panel d. The results shown here were also reproduced by replacing the temperature with the Planck function. Because the temperature is somewhat more intuitive than the Planck function we only show the results for the temperature.

\subsection{Statistical influence of horizontal radiative transfer}\label{sec:hdfe23_statistical}
%%%%%%%%%%%%%%%%%%%%%%%%%%%%%%%%%%%%%%%%%%%%%%%%%%%%%%%%%%%%%%%%%%%%%%%%
\begin{figure*}
\center{ \resizebox{0.7\hsize}{!}{\includegraphics{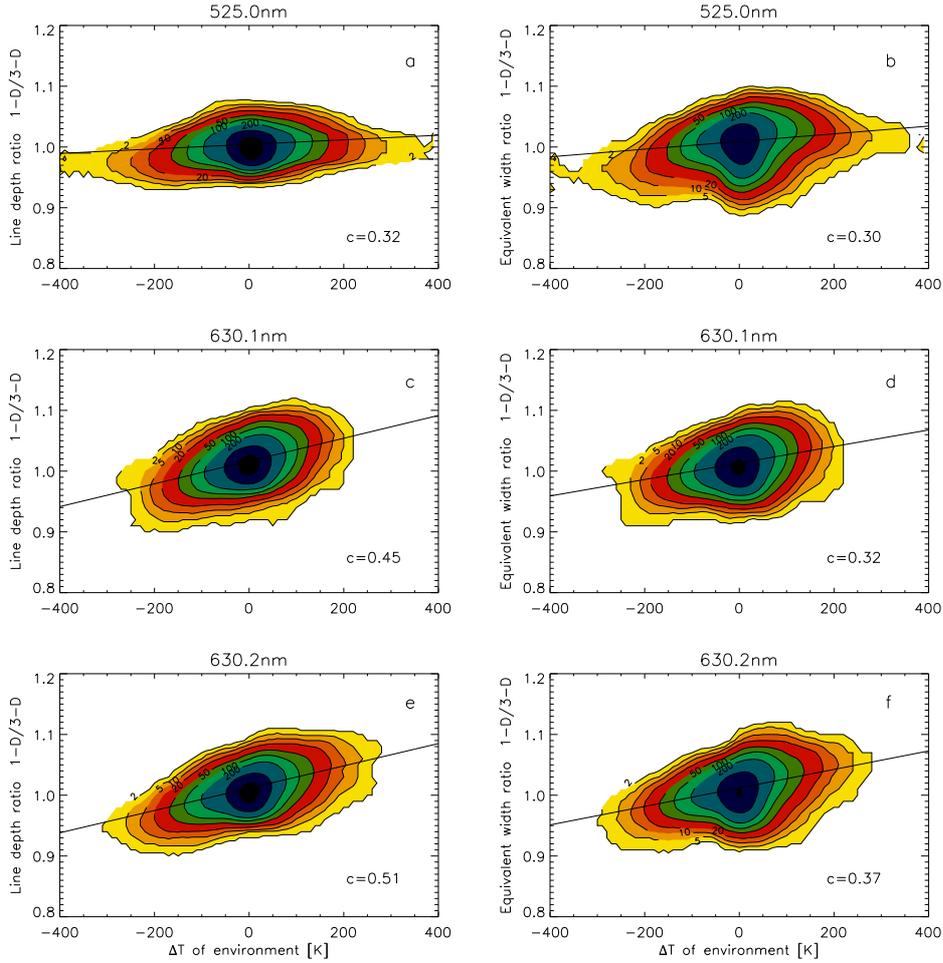}} }
 \caption{ 
Line depth ratio $D_{1-D NLTE}/D_{ \mbox{3-D} NLTE}$ (left panels) and analogous ratio of equivalent width (right panels) as a function of the temperature difference to the local environment for all $(x,y)$ positions of the hydrodynamic data cube. The point density is indicated by the colored shading in a logarithmic scale (yellow: 2, orange: 5, dark orange:10, red:20, etc. up to the innermost area in black with 1000+). The unit area was defined by subdividing the temperature axis into sections of $10$ K and that of the EW ratio into sections $0.01$. Regression lines are over-plotted and the corresponding correlation coefficients are given in the lower right corner of each frame.
} 
 \label{fig:scatter_dT}
\end{figure*}

\begin{table}
\caption{
Slopes (in \%) of the linear regression lines in Fig.~\ref{fig:scatter_dT}.
%temperature
% 1.0043682 1.0090620 1.0165838 1.0133582 1.0115854 1.0119878
% 0.37397369 0.63131562 1.8746017 1.3580882 1.8353814 1.5214741
}
\begin{tabular}{lrr}
\hline
 Line & Line depth & equivalent width \\
\hline
$525.02$~nm & 0.374 & 0.631 \\
$630.15$~nm & 1.875 & 1.358 \\
$630.25$~nm & 1.835 & 1.521 \\
\hline
\end{tabular}
\label{tab:sensitivity_T}
\end{table}

In Sect. \ref{sec:hdfe23_results_origin} we quantified the differences between \mbox{1-D} NLTE and \mbox{3-D} NLTE discussed in the previous section, for a few sample points only, which demonstrate the basic effect. Here we consider all points of the whole atmosphere and all four computed spectral lines at the heights of line formation. In Fig.~\ref{fig:scatter_dT} we plot the two ratios (left panels: line depths $D_{1-D}/D_{3-D}$; right panels: equivalent widths $W_{\lambda,1-D}/W_{\lambda,1-D}$) against the average temperature difference $\Delta T^{Env}(x,y)$ between the line-forming region and its environment for all points $(x,y)$ in the atmospheric model, whereas $\Delta T^{Env}(x,y)$ is defined as the average of all $\Delta T^{Env}_{x,y,z}$ in the formation height range $F_{line}(x,y)$ (see Eq. \ref{eq:lineformingregion}). The result for \ion{Fe}{i} $524.71$~nm is not plotted since it is nearly the same as for $525.02$~nm.

Even with our very crude definition of the environmental temperature, a clear, almost linear trend can be seen: The hotter the environment the weaker the \mbox{3-D} NLTE line profile and vice versa. This corresponds exactly to our results of the previous section.

For every line we may now determine the specific slopes of the line depth or $W_{\lambda}$ ratios as a function of the temperature difference of the environment. The slope is different for every line in correspondence to their sensitivity to NLTE and horizontal RT effects. Table \ref{tab:sensitivity_T} gives the slopes of the linear regression in \% per $100$ K temperature difference. E.g. the line depth of the $630.25$~nm line in \mbox{1-D} NLTE must be increased by $1.835$\% (the equivalent width by $1.521$\%) for every $100$ K which the "environment" is cooler than the line-forming region to obtain the \mbox{3-D} NLTE value. It can be seen that both the \mbox{3-D} line depth and $W_{\lambda}$ decrease by approximately 1 to 2\% per $100$ K temperature difference in the selected lines. The zero-crossing of the regression lines is at $1.01$ for both quantities, i.e. both, the \mbox{1-D} NLTE line depths and the equivalent widths are on average $1$\% larger than the corresponding \mbox{3-D} NLTE value.

An improved definition of the environment could reduce the scatter in the plots. A test with other distances $\Delta x, \Delta y$ of the four pixels confirmed that the selected distance of approximately $100$~km is appropriate, but the inclusion of more pixels, possibly weighted according to their optical distance, is likely to give better correlations. However, the more sophisticated the method to determine the neighborhood, the higher are the computational costs. 

Could one use such relationships as simple and still reasonably effective first order corrections of \mbox{1-D} NLTE values which could allow one to bypass full-blown \mbox{3-D} computations? Since the temperature of neighboring pixels is not known a priori in inversions, we replaced our temperature measure of the environment by the surface residual intensity of the line in the same environment. The resulting scatter plot (not shown) shows a very similar behavior as Fig.~\ref{fig:scatter_dT}. A possible reason is that if we assume that the lines are formed at least close to LTE, the residual intensity in the neighborhood is also a weighted measure of the temperatures in that neighborhood and therefore a similar connection between the residual intensity of the environment and the relative strength of the \mbox{3-D} line profile must exist.

% intensity
% 1.0042986 1.0089445 1.0165664 1.0133456 1.0118133 1.0121767
% 2.8559488e+11 5.1000472e+11 2.0813935e+12 1.6626431e+12 1.1902803e+12 1.1126243e+12

\section{Discussion}\label{sec:hdfe23_discussion}

In the previous section we presented the influence of the inclusion of NLTE effects and horizontal RT on the profiles of spectral lines. We obtained these results by applying three different calculation methods (namely \mbox{3-D} NLTE, \mbox{1-D} NLTE, and LTE) on a realistic \mbox{3-D} model of the solar photosphere. We obtained partly substantial differences when spatially resolved quantities such as the line depth, equivalent width, etc. were considered. Spatially averaged values, however, are only rather weakly affected by horizontal RT, and although NLTE effects may have some influence, they remain small in most cases. An important exception is the RMS contrast, a global quantity that is strongly influenced by horizontal RT. 

The influence of the inclusion of NLTE effects when computing spectral lines in \mbox{3-D} HD models has been investigated in detail by \citet{shchukinatrujillobueno2001}. Here, we therefore concentrate mainly on discussing the influence of horizontal RT.

The main reason for differences between calculations considering or neglecting horizontal RT can be generally ascribed to the effect already discussed by \citet{stenholmstenflo1977}. They showed that the influx of radiation from a hot environment leads to a weakening of \ion{Fe}{i} spectral lines (later confirmed for more realistic iron model atoms and flux-tube geometries by \citet{brulsvdluehe2001} and \citet{holzreutersolanki2012}. However, in granulation \citep[and in some flux-tube models, see][]{holzreutersolanki2012} a full explanation requires the addition that the \mbox{3-D} NLTE line profile may not only be weakened by the irradiation from a hot environment, but may be strengthened if the environment is sufficiently cold. Furthermore, as the formation height of the considered lines lays slightly above the granular-intergranular temperature inversion, the over-all effect may be weaker in these lines because the contributions from both below as well as above the temperature inversion height may cancel each other.

\subsection{Interpretation of contrasts}\label{sec:hdfe23_discussion_contrasts}
%%%%%%%%%%%%%%%%%%%%%%%%%%%%%%%%%%%%%%%%%%%%%%%%%%%%%%%%%%%%%%%%%%%%%%%%

%\begin{figure}
%\center{ \resizebox{0.85\hsize}{!}{\includegraphics{./RMS_CLV_6301_lamMin.ps}} }
% \caption{ 
%Influence of a (hypothetical observational) filter width (Gaussian) on the residual intensity contrasts. The %residual intensity contrasts are given for two positions on the solar disc, $\mu = 0.32$ (red) and $\mu = 1.00$ %(black), as well as our three calculation methods (asterisks: LTE; plus signs: \mbox{1-D} NLTE; triangles: \mbox{3-D} NLTE). 
%} 
% \label{fig:filter_Imin}
%\end{figure}

\begin{figure}
\center{ \resizebox{0.85\hsize}{!}{\includegraphics{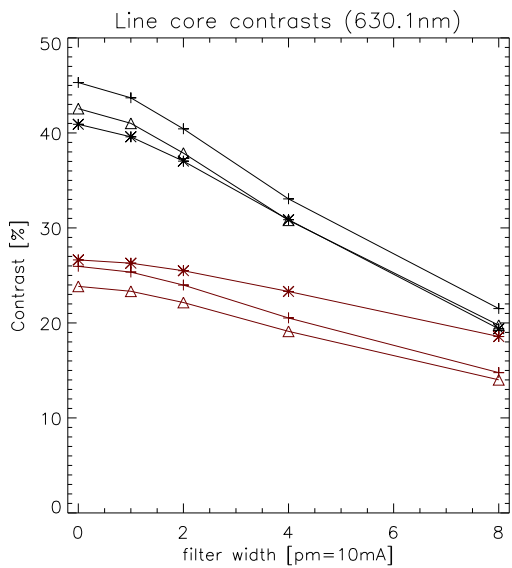}} }
 \caption{ 
%As Fig.~\ref{fig:filter_Imin} but for the contrasts at the nominal line core wavelength.
Influence of a (hypothetical observational) filter width (Gaussian) on the intensity contrasts at the nominal line core wavelength. The contrasts are given for two positions on the solar disc, $\mu = 0.32$ (red) and $\mu = 1.00$ (black), as well as our three calculation methods (asterisks: LTE; plus signs: \mbox{1-D} NLTE; triangles: \mbox{3-D} NLTE). 
} 
 \label{fig:filter_I0}
\end{figure}

In Sect. \ref{sec:hdfe23_result_contrast} we found that the different levels of realism in the RT may produce very different RMS contrasts within the spectral lines. Horizontal RT reduces contrasts in a similarly strong way as NLTE effects do. The two effects stack in such a way that differences of up to a factor of $2$ between LTE and \mbox{3-D} NLTE and up to a factor of $1.5$ between \mbox{1-D} NLTE and \mbox{3-D} NLTE may arise. The largest effect of horizontal RT is found in the residual intensity contrast owing to the low absolute value of the intensity. The occurrence of such strong discrepancies between different methods has an impact on the interpretation of observed contrasts, recorded in the cores of spectral lines or in spectral regions with a dense coverage by spectral lines, such as found in the UV and observed by the Sunrise filter imager \citep{Gandorferetal2011}. 
 
Spectral filters invariably have a finite width. In order to compare our calculated contrasts with those of such an observation, we convolved our calculated line profiles with Gaussian filters of different widths. Figure.\ref{fig:filter_I0} presents the contrast after a filter tuned to the nominal line core wavelength as a function of the filter width. Black and red curves are for $\mu=1.0$ and $\mu=0.32$, respectively. 

A filter as narrow as $4$~pm reduces contrast considerably and filter widths of $8$~pm almost halve the contrast values. However, even for filters of this width a significant difference in contrast is found between LTE and \mbox{3-D} NLTE amounting to a factor of $1.3$ at $\mu=0.32$ (see Fig.~\ref{fig:filter_I0}). At double this filter width (i.e. $16$~pm) the recorded radiation is dominated by the continuum, so that the contrast becomes close to the continuum contrast at this wavelength \citep[see][]{danilovicetal2008}. Since the continuum is formed in LTE, the differences between the results of the three levels of sophistication become accordingly small.

We wish to draw attention to the fact that in LTE the CLV of the contrast changes rather dramatically as the filter width increases: The LTE contrast for larger filter width is almost the same for both $\mu$ so that CLV of the contrast is strongly distorted. In NLTE, the ratio of the contrast at $\mu=1.0$ to that at $\mu=0.32$ is reduced (from 1.8 for no spectral smoothing to 1.4 at $8$~pm filter width), but remains sizable.

% der folgende Absatz ist wichtig aber wenig aufschlussreich. Was man tun müsste ist die 3 Resultate der drei Methoden räumlich verschmieren (2D gauss) und schauen wie sich der Kontrast mit Verschmierung verhält. Ohne eine solche Analyse würde ich den absatz deutlich kürzen.
The contrast also depends significantly on the spatial resolution. In the past, the rule that the finer the resolution the larger the contrasts was true \citep{mathewetal2009,wedemeyervanderVoort2009}. We cannot make a testimony on whether we have reached the end of this development or not, as our HD atmosphere has only a finite resolution ($21$~km per pixel) too, but horizontal RT at least has some smoothing effect on such small scales so that we can speculate that the increase in contrast with resolution could flatten somewhat. However, we found in Paper~I \citep[see also][]{brulsvdluehe2001} that very small features with large contrast may persist in magnetic atmospheres so that we must assume that we can expect some new findings when going to better resolution. We intend to expand our series of investigations to MHD atmospheres with resolutions as high as $5$~km per pixel. 

Because our resolution of approximately $20$~km is far better than those of the best observations available today and because we do not experience any smoothing due to the point-spread function of the optical system our contrasts are much larger than observed values.

To summarize this section, we note that horizontal RT has a strong influence on the determination of contrasts and neglecting it may lead to severe errors in the interpretation of observational contrasts measured in spectral lines. Spectral filters with a width in excess of $4$~pm ($40 $m$\AA$), as often used for high-resolution imaging, have a strong influence on the contrasts observed in the cores of isolated medium strong spectral lines. The CLV of the contrast may be strongly affected, whereby the effect is larger for LTE computations than in NLTE, at least for the lines considered here.

\subsection{The influence of horizontal radiative transfer on temperature determination}\label{sec:hdfe23_discussion_Tdetermination}
%%%%%%%%%%%%%%%%%%%%%%%%%%%%%%%%%%%%%%%%%%%%%%%%%%%%%%%%%%%%%%%%%%%%%%%%
In their extensive investigation, \citet{shchukinatrujillobueno2001} confirmed that for spatially resolved observations of \ion{Fe}{i} spectral lines neglecting (1-D) NLTE effects may influence the determination of the temperature. Such results had earlier been found by \citet{ruttenkostik1982}, \citet{shchukinaetal1990}, \citet[][ for magnetic features]{solankisteenbock1988} and others. They found, that the determined temperature obtained from the $630.15$~nm line may be $100-200$ K too large in granules and too small by approximately the same amount in intergranular regions. The profiles in Fig.~\ref{fig:cut_profiles} of the slightly weaker $630.25$~nm line, reveal that the differences between the \mbox{3-D} and \mbox{1-D} NLTE profiles can be of the same order of magnitude as those between the \mbox{1-D} NLTE and LTE profiles. Hence, neglecting horizontal RT may produce errors of similar size, at least at some locations.

We find the strongest effects of horizontal RT e.g. on the line depth or EW --- as expected --- at the edge of the granules, where strong horizontal temperature and density variations at heights relevant to line formation are present. These locations only partly overlap with the locations found by \citet{shchukinatrujillobueno2001} where NLTE is important (i.e. their "granular region"). Therefore, the differences due to horizontal RT can amplify or diminish the error of neglecting NLTE in the temperature determination, as \mbox{1-D} NLTE effects are mainly due to the \emph{vertical} structure of the atmosphere, while the effects of horizontal (3-D) RT are mainly determined by the \emph{horizontal} structure. 

If we consider the sample profiles in Fig.~\ref{fig:cut_profiles}, we would clearly assign the profile d to be in an intergranular lane (refer also to Fig.~\ref{fig:tv_atmos_blowup}). If -- according to the lower left panel of Fig. 14 of \citet{shchukinatrujillobueno2001}, which corresponds closely to the case in our Fig.~\ref{fig:cut_profiles}d -- applying LTE leads to an underestimate of the temperature by $100-200$ K at such a location, one can conclude that the LTE temperature estimate at this point would be $200-400$ K lower than the \mbox{3-D} NLTE value because the difference between the \mbox{1-D} and the \mbox{3-D} NLTE profiles is of the same magnitude as that between the LTE and \mbox{1-D} NLTE profiles. An analogous interpretation, but with reversed sign can be applied to panels a and b. At the very border of a granule, the LTE temperature estimate in the \ion{Fe}{i} $630.15$~nm line could be $200-400$ K too high. 

An example of horizontal RT reducing the error in temperature determination is illustrated in Fig.~\ref{fig:cut_profiles}f, whereas Figs.~\ref{fig:cut_profiles}c and e depict profiles that do not experience any change due to horizontal RT. Note also that the formation height of the considered lines lays above the granular-intergranular temperature inversion, if a weaker line would be considered, the effects of horizontal RT could have the reversed sign.

\subsection{Yet another word on the iron abundance}\label{sec:hdfe23_discussion_abundance}
%%%%%%%%%%%%%%%%%%%%%%%%%%%%%%%%%%%%%%%%%%%%%%%%%%%%%%%%%%%%%%%%%%%%%%%%
How important is horizontal RT for abundance determinations? The strong differences between \mbox{1-D} and \mbox{3-D} NLTE in Fig.~\ref{fig:histo_ratio_Aeqw} may look like an alarming result for those who determine abundances by determining equivalent widths. Although differences of up to $10$\% at specific locations, the average ratio (vertical bars in Fig.~\ref{fig:histo_ratio_Aeqw}) between the \mbox{1-D} and \mbox{3-D} value is at slightly more than $1.01$ corresponding to a difference of approximately $1$\% in EW. As the curves of growth of both the $630$~nm lines reaches its lowest inclination close to the actual abundance values, a small but relevant change in the iron abundance of approximately $0.012$ dex may result. The same result is obtained if one considers the EW of the two spatially averaged profiles instead of the spatial average of the individual EW's. The use of weaker lines that are less susceptible to NLTE effects may reduce the effect of neglecting horizontal RT. 

Our finding is not really surprising as we consider a field-free atmosphere, where no systematic difference in density between hot and cold environments can be found. Only in an evacuated magnetic element where the hot walls can irradiate into the flux element, can the line weakening systematically dominate over the line strengthening which takes place only in a very small area in the optically much thicker walls. 

%dex 
%a logarithmic unit being used in astronomy. Originally, dex(x) = 10^x. But the notation is now being used after the exponent in expressions such as -.043 dex, meaning 10-.043. Thus 1 dex equals a factor of 10. The name "dex" is a contraction of "decimal exponent." 
%
% 7.50 Steigung-> 0.02 dex entspricht Faktor 1.0171 ie 1.7prozent in eqw.
% 1 Proz EW entspricht 0.0116 dex

% \mbox{3-D} 4.3515665e-10
%LTE 4.5567447e-10
%1-D 4.3971899e-10

\section{Conclusions}\label{sec:hdfe23_conclusions}
%%%%%%%%%%%%%%%%%%%%%%%%%%%%%%%%%%%%%%%%%%%%%%%%%%%%%%%%%%%%%%%%%%%%%%%%
This paper describes a continuation of our investigation of Paper~I on the influence of horizontal radiative transfer (RT) in NLTE on diagnostically important \ion{Fe}{i} lines. The main difference to Paper~I lies in the fact that we replaced the simple flux tube model atmosphere used there by a snapshot of a realistic \mbox{3-D} radiation hydrodynamic simulation (with $B=0$) run of the MURAM code \citep{voegleretal2005}. We kept the original resolution of $288 \times 288$ pixels in the $xy$ plane which allowed us to investigate horizontal transfer in high spatial resolution (horizontal grid scale $\approx 20$~km) spectra. 

Our results confirm the line weakening influence of radiation streaming in from hot surroundings, as described by \citet{stenholmstenflo1977} and in Paper~I. In the realistic field-free atmosphere considered here, we also find the inverse effect of line strengthening due to weaker irradiation from colder environments. In normal granulation, both effects occur almost equally often. This leads to a surprisingly good agreement of the spatially averaged profiles calculated in \mbox{1-D} NLTE with those computed in \mbox{3-D} NLTE. The difference to profiles computed in LTE can be quite significant, however. Furthermore, the fact that the lines considered here are formed in heights only slightly above the granular-intergranular temperature inversion may lead to a partial cancellation of the effects of horizontal RT because contributions from heights below the temperature inversion typically have a different sign than those from above it. Lines built below the temperature inversion height might have an augmented sensitivity to horizontal RT.

\citet{shchukinatrujillobueno2001} found that NLTE effects may not be negligible even if averaged profiles calculated in LTE or NLTE agree well. Our computations show that the same can be said for the influence of horizontal RT, where significant differences to \mbox{1-D} NLTE in the individual profiles can be found, although the averaged profiles coincide almost completely. Equivalent widths and line depths calculated in true \mbox{3-D} NLTE differ mainly at spatial locations close to strong horizontal temperature gradients -- as found at the boundaries of granules. These parameters in \mbox{3-D} NLTE may differ by up to 20\% from their LTE counterparts and by up to 10\% from the values obtained with \mbox{1-D} NLTE. The average difference between the \mbox{1-D} and \mbox{3-D} NLTE equivalent widths as well as the difference of the equivalent widths obtained by spatially averaged profiles amounts to slightly more than $1$\% for our atmosphere and would result in a reduction of the iron abundance of approximately $0.012$ dex. 

The largest influence of horizontal RT has been found in the RMS contrast of the line core intensity. While NLTE reduces the residual line contrast by approximately one third against the LTE value, we found that horizontal RT reduces the \mbox{1-D} NLTE contrast by another 20 to 30\%, resulting in a contrast that is only roughly half of that obtained in LTE. This is valid for all distances from the limb. Contrasts measured using filters centered on the nominal wavelength of the line core depend strongly on the width of the filter. However, at all filter widths the CLV of the contrast resulting from LTE and \mbox{3-D} NLTE computations is different.

We have shown in detail how the temperature difference between the line-forming region and its environment systematically weakens (if the environment is hotter) or strengthens (if the environment is colder) the profiles calculated in \mbox{3-D} NLTE. A clear statistical correlation exists between the line weakening (strengthening) as a function of the temperature of the environment. A quantitative investigation shows that the weakening or strengthening of the $630$~nm lines by horizontal RT is on the order of $1-2$\% of line depth or equivalent width per $100$ K temperature difference to the horizontal environment of the line-forming region. A similar correlation exists also with the line intensity in the (same) neighborhood, as the line intensities are also dominated by the local temperatures as long as the line does not depart too much from LTE, which is the case for the considered lines. 

Our results are of particular importance for the inversion of high resolution observations. If we invert an observation with resolution below $100$~km in \mbox{1-D} NLTE, then our calculated profiles will locally deviate by up to several percent in line depth or equivalent width from \mbox{3-D} NLTE. As a consequence, temperatures obtained by such an inversion could also be wrong by several ten to a few hundred Kelvin. The aforementioned correlation between the relative environmental temperature and the 1-D$/$ \mbox{3-D} line depth ratio could in principle be used to coarsely correct inversions done in \mbox{1-D} NLTE for the influence of horizontal RT. The errors in temperature between LTE and \mbox{1-D} NLTE also amount to $100-200$ K \citep[][]{shchukinatrujillobueno2001}, so that locally errors up to $300-400$ K can occur between LTE and \mbox{3-D} NLTE. However, the errors due to neglecting (1-D) NLTE and due to neglecting horizontal RT can also cancel each other as enhance each other, depending on the properties of the surroundings. 

The next step in this investigation on important iron lines will be the use of \mbox{3-D} NLTE simulations including a magnetic field. The combined influence of horizontal RT and Zeeman effect is likely to lead us to new insights about relevant RT processes in the solar photosphere. In particular, we expect to learn how well a simple flux tube model (Paper~I) compares with realistic MHD simulations, with regard to line profile changes due to horizontal RT.

%%%%%%%%%%%%%%%%%%%%%%%%%%%%%%%%%%%%%%%%%%%%%%%%%%%%%%%%%%%%%%%%%%%%%%%%%
\begin{acknowledgements} 
We are grateful to Manfred Sch\"ussler for providing the data cubes resulting from the MURAM simulation run, as well as Han Uitenbroek for making his excellent full Stokes NLTE radiative transfer code available. Jo Bruls has provided the iron model atom and has given us many insights into NLTE effects. R.~H. appreciates the flexibility of Prof. Dr. Norbert Dillier concerning the working hours at the University Hospital of Zurich. This work has been partially supported by WCU grant No. R31-10016 funded by the Korean Ministry of Education, Science and Technology. And we thank the referee for his help to improve the manuscript. 
\end{acknowledgements}

\bibliographystyle{aa}
\bibliography{journals,holzreuter}

%%%%%%%%%%%%%%%%%%%%%%%%%%%%%%%%%%%%%%%%%%%%%%%%%%%%%%%%%%%%%%%%%%%%%%%%
\end{document}